\documentstyle[10pt,epsf,twoside]{article}

\newcommand{\beq}{\begin{equation}}
\newcommand{\eeq}{\end{equation}}
\newcommand{\eq}[1]{Eq.(\ref{#1})}
\newcommand{\eqo}[1]{(\ref{#1})}
\newcommand{\eqs}[1]{Eqs.(\ref{#1})}

\title{Some possibilities for laboratory
searches for variations of fundamental constants}

\author {Savely G. Karshenboim\thanks{E-mail: sek@mpq.mpg.de; ksg@hm.csa.ru}\\
\medskip
 {}\\
 D. I.  Mendeleev Institute for Metrology (VNIIM),\\  St. Petersburg 198005,
 Russia\\
 {}\\
 and\\
 {}\\
 Max-Planck-Institut f\"ur Quantenoptik,\\
 85748 Garching, Germany\thanks{The summer address}
}

\begin{document}

\maketitle

\begin{abstract}
We consider different options for the search for possible
variations of the fundamental constants. We give a brief overview
of the results obtained with several methods. We discuss their
advantages and disadvantages with respect to simultaneous
variations of all constants in both time and space in the range
$10^{8}-10^{10}$ yr. We also suggest a few possibilities for the
laboratory search. Particularly, we propose some experiments with
the hyperfine structure of hydrogen, deuterium and ytterbium--171
and of some atoms with a small magnetic moment. Other suggestions
are for some measurements of the fine structure associated with
the ground state. Special attention is paid to the interpretation
of the {\em hfs\/} measurements in terms of variations of the
fundamental constants.
\end{abstract}

\section{Introduction}

There is no physical reason to expect that the ``fundamental
constants'' are really constant quantities. Indeed, their
variations have to have a cosmological scale. Since the publishing
of a famous Dirac paper \cite{dirac} a number of different
hypotheses on their possible variations have been suggested as
well as a number of trial to search for these. A review of old
models and searches could be found e. g. in Ref. \cite{dyson}. We
live in an expanding universe and an accepted contemporary picture
of the history of our universe assumes that there were a few phase
transitions with spontaneous breaking of some symmetries
during the early stages of evolution (inflation model)
\cite{inflation}.

We consider here only atomic and nuclear properties and
the variations of the fundamental constants that can be determined
from changes in such properties.
We do not discuss variation of the gravitational constant (a
review on that can be found in Refs. \cite{dyson,sisterna}). There
are two reasons for that. First of all, in looking for variations
of nuclear and atomic properties (magnetic moments, masses, decay
rates) it is not possible to consider strong, weak and
electromagnetic effects independently. Next, from a theoretical
point of view, we have to expect that the variations of strong,
weak and electromagnetic coupling constants are strongly
correlated. In contrast to that, any investigation for possible
variations of the gravitational interaction can be done
separately.

\subsection{Variation of the constants and atomic spectroscopy}

 Any search for a possible variation of the
fundamental constants can be actually performed measuring some
 atomic, molecular and nuclear properties and it is necessary
to discuss relations between them.
There are three very different kinds of
atomic and molecular transitions, available for precision spectroscopy:
\begin{itemize}
\item {\em Gross structure\/} is associated
with transitions between levels with different values of the
principal number $n$. They have a scaling behaviour as $\alpha^2\,
m_e\,c^2/h$ or $Ry$. It has to be remembered that for any atom
with two or more electrons, one has to introduce an effective
quantum number $n^*$, which is a function of the orbital momentum
($l$). The nonrelativistic behaviour $\alpha^2\, m_e\,c^2/h$ is
actually associated with $n^*(l)$ and, hence, in
non-hydrogen-like atoms, transitions between levels with
different value of $l$ are also a part of the gross structure.
\item Atomic {\em fine structure\/}, or separations of levels with the same
$n^*(l)$ and different spin-orbit coupling (e. g. different $j$
or different sum of spin of valence electrons), has a scaling
behaviour $\alpha^4\, m\,c^2/h$ or $\alpha^2\,Ry$. That is part of
relativistic corrections, which have the same order of magnitude.
\item Atomic
{\em hyperfine structure\/} is a splitting due to a magnetic moment of
the nucleus ($\mu$) and it is proportional to
$\alpha^4\, m\,c^2/h \,(\mu/\mu_{\sc b})$ or $\alpha^2\,(\mu/\mu_{\sc b})\,Ry$.
\end{itemize}

Comparing frequencies with different scaling behaviour, one can
look for variations of the fundamental constants and nuclear
magnetic moments. An important part of those comparisons is the
so-called absolute frequency measurements. An absolute
measurement of a frequency is actually its comparison to the {\em
hfs\/} of Cs, because of the definition of {\em second\/} via the
Cs {\em hfs\/} interval
\begin{equation} \label{Cshfs}
\nu_{\rm hfs}(^{133}{\rm Cs}) = 9\,9192\,631.770 ~{\rm kHz}\quad
{\rm (exactly)}\,.
\end{equation}

Another possibility is to study some corrections. E. g. the study
of relativistic corrections to the atomic gross structure or {\em
hfs\/} can yield to detection of possible variations of the fine
structure constants, while investigations of the
electronic-vibrational-rotational lines can give limitations for
the variation of the proton-to-nuclear mass ratio. Due to the
importance of the value in \eq{Cshfs} let us to discuss
corrections to the {\em hfs\/} in some detail. The value of the
hyperfine splitting in an atomic state can be presented in the
form
\beq \label{theohfs}
\nu_{\rm hfs} = c_0 \, \alpha^2 \, \frac{\mu}{\mu_{\sc b}}
\,{\rm Ry}\,F_{\rm rel}(\alpha)\,,
\eeq
where $\mu$ stands for the nuclear magnetic moment, $c_0$ is a
constant specific for the atomic state and all dependence on
$\alpha$ is contained in a function $F_{\rm rel}(\alpha)$, which
is also specific for the state. The function is associated with
the relativistic corrections and $F_{\rm rel}(0)=1$. The
corrections are more important for high $Z$ and in the case of low
$Z$ (e. g. for the hydrogen {\em hfs\/}) they are negligible.

\subsection{Variable particle, atomic and nuclear properties \label{scal}}

There are three possibilities for the variation of fundamental
properties, which can be studied within spectroscopic methods:
\begin{itemize}
\item effects due to electromagnetic interactions and value of electromagnetic
coupling constant $\alpha$ (study of $1s-2s$,
comparison of the gross and fine structure, study of relativistic corrections);
\item effects of quark-quark and quark-gluon strong interactions and
nucleon properties like $\mu_p$, $\mu_n$, $m_p$ etc
(study of the {\em hfs\/} of hydrogen and light nuclei, or molecular
electronic-vibrational-rotational lines);
\item nuclear properties (moments and masses)
due to the structure of heavy nuclei. Some of those properties
(like magnetic dipole and electric quadrupole moments) can be
studied by means of atomic and molecular spectroscopy. Nuclear
decay rates and scattering cross sections of nuclear collisions
can be investigated by other methods.
\end{itemize}

We mainly consider here applications of the precision spectroscopy.
Experiments with the {\em hfs\/} of heavy atoms (e. g. Cs, Rb,
Yb$^+$ or Hg$^+$) cannot have any clear interpretation: any of the
values are proportional to a non-relativistic matrix element, but
with significant relativistic corrections (see \eq{theohfs}). The
non-relativistic value is proportional to a magnetic moment of
the nucleus, which includes the moments of one or two valence
protons and neutrons and some contribution of an internal nuclear
motion. The last is not a pure effect of the strong interactions
and an influence of electromagnetic interactions can be estimated
from the proton-neutron asymmetry.

If we keep in mind a general picture, we have to expect a kind of
``grand unification'' theory and there has to be some direct
relations between coupling constants for weak, electromagnetic and
strong interactions (see e. g. \cite{guts,inflation}). Hence,  all
fundamental constants (i. e. $\alpha$, $\alpha_{\sc s}$ and
$\alpha_{\sc w}$) are expected to vary within about
the same rate. Indeed, different atomic, molecular or nuclear
properties can have quite different sensitivities to variations of
the coupling constants.

It is necessary to emphasise that it is incorrect to think that
some values like proton-electron mass ratio or nuclear
$g$-factor are expected to be relatively constant, while the
electromagnetic coupling constant $\alpha$ varies, as is expected
in the interpretation of some papers. The bare electron mass is a
result of the interaction with the Higgs field \cite{weak} and it
has significant quantum electrodynamical corrections due to the
renormalization. The bare $u$- and $d$-quark masses also appear
from the Standard model as a result of interaction with the Higgs
sector and are a few Mev. However, the actual masses of the proton
and neutron are determined by the masses of the constituent
quarks, which are about 300 Mev and those values are completely a
result of the dynamic effects of the QCD in a strong coupling
range. The same is true with the proton and neutron magnetic
moments. Such a value, appearing as a result of the strong
interactions, is definitely a function of the strong coupling
constant, which is expected to vary together with the
electromagnetic and weak coupling constants.

Let us consider the electron mass in more detail. A standard
interpretation of quantum electrodynamics (QED) is that it has to
be possible to explain all low-energy (with respect to the Planck
mass $M_{\sc p}$) physical phenomena, using only few effective
low-energy parameters (like renormalized electron mass and charge
etc). The origin of those values is not important. However, in the
case of the search for variations of the constants it is necessary
to study the origin of the low-energy parameters, like the
electron mass. The bare electron mass $m_0$ is to be renormalized
due to the QED effects. In the one-loop approximation on can find
\beq \label{emass}
m_e = m_0 \left[1+\frac{3}{4}\,\frac{\alpha}{\pi}\,
\ln\left(\frac{\Lambda}{m_0}\right)^2\right]\,,
\eeq
where the $\Lambda$ is an effective cut-off of the ultraviolet
divergence. If the cut-off is associated with the Planck scale
$\Lambda = M_{\sc p}\sim 1.2\cdot 10^{19}~{\rm Gev} \sim 2.4\cdot
10^{21}~ m_e$, the logarithmic correction is about 20\% of the
leading term. One can see that, even in the case of a constant
value of $m_0$, the actual electron mass $m_e$ has to vary with
$\alpha$.

From a theoretical point of view we can rather expect a primary
variation of some parameters which are not visible directly in our
low-energy world and this forces some secondary variations of the
coupling constants. E. g. in the inflatory universe (see Ref.
\cite{inflation} for detail), some effective Higgs potentials
depend on the average temperature, which is a function of the
time. That have led to some phase transition in the past, when the
hot universe was cooling. We think that even in our ``cool''
universe we can expect some slow time variation of those Higgs
potentials and perhaps their long-scale space variation. The
variations of the constants is to be a direct consequence of that.

Another and more sophisticated idea \cite{marciano} was proposed
due to the so-called Kaluza-Klein theories, which are associated
with a world of $4+N$ dimensions. In contrast to ``our'' 4
dimensions, the extra spatial $N$ dimensions form a compact
manifold with a radius $R_{\sc kk}$ about Planck length ($\sim
10^{-33}$ cm). While our four dimensional universe is expanding,
Marciano suggested that value of $R_{\sc kk}$ is also not
constant. However, in the Kaluza-Klein theory this value is
associated with the coupling constants of our world.

So, we expect that the variation of different coupling constants
has to have the same scale. However, it is necessary to remember
that due to the strong coupling any effective parameters coming
from the strong interactions, can have a kind of random variation.
We cannot often know which quantity is varied more (or less)
rapidly than the fundamental constants. That is why we have to try
with several ways as different as possible.

No model on a possible dependence of the values of the fundamental
constants has not been assumed in our paper. However, we must
underline, that we consider a picture of simultaneous variations
of all coupling constants, which particularly determine all
property of particles, nuclei, atoms and molecules.

We have not
yet specified a term ``variations''. To our mind there are actually two
possibilities for those:
\begin{itemize}
\item A time and/or space variation over all the universe with a cosmological scale
($T\sim 10^{10}$ yr and $L\sim c \cdot 10^{10}$ yr).
\item Time and/or space fluctuations over some less, but significantly
cosmological, scale ($T\sim 10^8-10^9$ yr and $L\sim c \cdot
(10^8-10^9)$ yr).
\end{itemize}
Such a fluctuation of the gravitation interaction was considered
\cite{hill,morikawa,crittenden,sudarsky,sisterna94,salgado} due to
a periodicity in the galaxies distributions in the direction of
the Galactic north and south poles \cite{broadhurst}. The
fluctuations of gravitation is also one of the explanations of
possible variations to the solar year \cite{sisterna94}.

The variation can be induced e. g. by cooling of the universe and
a variation of some effective Higgs potentials. We have no a
priori estimation for the speed of the variations. We expect the
cosmological time and space scale ($T$ and $L$), but we have no
idea on the amplitude of the possible variations. The expected
amplitude particularly depends on our assumption if we expect some
kind of a primary direct variation of the constants, or their
variations are a consequence of variations of some other
parameters. E. g. let us pretend that the primary variation is due
to the compactification radius $R_{\sc kk}$ and there is no
variation of the coupling constants on the unperturbed level.
Nevertheless, the variations have to appear due to the
renormalization and the time-dependence is to be of the relative
order $\alpha \ln{R_{\sc kk}(t)}$. Particularly, following QED one
can find in the one-loop approximation an expression for the fine
structure constant
\beq \label{alpha}
\alpha = \frac{\alpha_0}{1-\frac{1}{3}\,\frac{\alpha_0}{\pi}\,
\sum_{j}{c_j}\,q_j^2\,
\ln\bigl(\Lambda/m_j\bigr)^2}~,
\eeq
where $\Lambda\simeq \hbar/cR_{\sc kk}$, the sum is over all fundamental
charged particles (leptons, quarks, $W$-bosons, Higgs particles
etc), $m_j$ stands for their masses, $q_j$ is for their charges.
The coefficient $c_j$ is dependent on their spin and particularly
for 1/2 it is equal to one. A significant variation of $R_{\sc
kk}$ (e. g. 1\%) can induce a variation of $\alpha$ only on a
level between $10^{-4}-10^{-5}$. The extra $\alpha$ and the
logarithm reduce the amplitude of the variation dramatically.
However, it is important to note, that the variation of the mass
in \eq{emass} with $\Lambda=R_{\sc kk}(t)$ and of the charge in
\eq{alpha} are to be of the same origin and of about the same
order. That example shows that the variation $\Delta
\alpha/\alpha$ can be small and that the mass variation $\Delta
m/m$ and the coupling constant variation can be of the same order
and vary simultaneously. Actually comparing \eqs{emass} and
\eqo{alpha} one can note that the variation of the mass can be
larger, smaller or of the same order of magnitude as the variation
of the coupling constant. Mass variations smaller or of the same
order as the coupling constant have been discussed above. The
larger mass variation can appear if we suggest that there is no
direct variation of the coupling constants, but the masses vary e.
g. due to some variation of the effective Higgs potential of the
Standard Model. The $\alpha$-variation has to appear from
\eq{alpha} because of the $m$-variation in the logarithm. We
conclude that a priori it is not feasible to dismiss the mass
variation and to consider only varying coupling constants.

\section{Non-laboratory search for variations of the fundamental constants}

\subsection{Geochemical data and nuclear properties}

Nuclear reactions (collisions, decay etc) often involve some
relatively small differences of large contributions. E. g. to
understand if any isotope is stable for a particular channel of
decay, one has to compare the initial and final binding energy. A
problem of the stability is a problem of this difference, which is
sometimes quite small. Relatively small shifts in particle masses
or coupling constants can make the decay of a stable isotope
possible, or disturb an allowed decay. The variation of the
coupling constant of strong, electromagnetic or weak interactions
can be weakly limited, but the geochemical estimations take
advantage of long term comparisons. The study of the abundance of
some isotopes allows one to make a comparisons over a geophysical
scale of $10^9$ yr.

\subsubsection{Geochemical data}

Some estimations of possible variations of the coupling constants
of strong, weak and electromagnetic interactions from geophysical,
or rather, geochemical data were performed in Refs.
\cite{wilkinson,peebles,dyson67,gold,peres,dyson,davies,lindner,barabash}
(see review in Refs. \cite{dyson,sisterna} for more references).
Some examinations also include data on the abundance of some
isotopes in the meteorites, and so the results are, in part,
astrochemical ones.

The typical limits are
\beq
\frac{1}{\alpha}\,\frac{\partial \alpha}{\partial t}<5\cdot 10^{-13}-
3\cdot10^{-15}
~{\rm yr}^{-1}\,,
\]
\[
\frac{1}{\alpha_{\sc s}}\,\frac{\partial \alpha_{\sc s}}
{\partial t}<5\cdot 10^{-11}-2\cdot 10^{-11}
~{\rm yr}^{-1}
\]
and
\[
\frac{1}{\alpha_{\sc w}}\,\frac{\partial \alpha_{\sc w}}
{\partial t}<10^{-10}
~{\rm yr}^{-1}\,.
\eeq
One problem with the interpretation of those data is the authors
assumed of Refs. \cite{dyson,dyson67,broulik,davies,lindner} that
only coupling constants vary, while the masses of the proton, the
neutron and the electron are constant. Conversely, we expect those
to vary as well, and some nuclear effects are sensitive to that
variation. Particularly, the $\beta$-decay must be very sensitive
to their difference
\[
m_n-m_p-m_e
\,.\]
We should remember that often the violation of the isotopic
invariance and particularly the small difference of the proton and
neutron masses
\beq
\frac{m_n-m_p}{m_p} \sim 0.14\%
\eeq
is associated with electromagnetic effects.

Next, it was assumed that it is possible to look for a variation
of some particular constant (e. g. $\alpha$) while others are
really constant. We cannot accept such an evaluation, but
nevertheless we would like to underline, that the nuclear property
can be very sensitive to a variation of the constants, because the
decay rates are strongly dependent on the transition energy, which
is actually a small difference of two larger energies of initial
and final states. Both include contributions of the
electromagnetic interactions and the difference can be quite
sensitive to these.

Another important problem is timing. Geochemical clocks are based
on the study of the abundance of some long-living isotopes and
others associated with them. A long lifetime is a result of a
small value of the transition either matrix element or energy.
Both, being small, are sensitive to the same variation of the
constants. To the best of our knowledge there have been no
discussions on a correlation between the clock and the variation.

\subsubsection{Geochemical data from Oklo reactor}

Shlyakhter \cite{shlyakhter} introduced two important elements in
the study of nuclear reactions. First, he pointed out that
laboratory investigations of nuclear property can also give
reasonable limitations (see section \ref{LabNucl}). His other idea
was due to the recently discovered Oklo Fossil reactor in Gabon
(West Africa). That is a natural fission reactor (see review in
Refs. \cite{oklo,petrov}), and conditions for its existence are
very narrow. Investigating those conditions and the local
abundance of different isotopes (particularly Sm) it is possible
to derive some limitations such that it had operated 1.7 billions
years ago for a period from 0.6 to 1.5 millions years. The limits
from Ref. \cite{shlyakhter} are
\beq \label{schl}
\frac{1}{\alpha}\,\frac{\partial \alpha}{\partial t}<1\cdot 10^{-17}
~{\rm yr}^{-1}\,,
\]
\[
\frac{1}{\alpha_{\sc s}}\,\frac{\partial \alpha_{\sc s}}
{\partial t}<1\cdot 10^{-18}
~{\rm yr}^{-1}
\]
and
\[
\frac{1}{\alpha_{\sc w}}\,\frac{\partial \alpha_{\sc w}}
{\partial t}<4\cdot 10^{-12}
~{\rm yr}^{-1}\,.
\eeq

One evidence of the operation of the nuclear reactor in the past
was isotope compositions of some elements like samarium, europium
and gadolinium. Some of their isotopes ($^{149}$Sm, $^{151}$Eu,
$^{155}$Gd and $^{157}$Gd) are strong neutron absorbers and they
have been found in very small quantities with respect to the
natural abundance. They have simply been burned by the flux of
thermal neutrons. A study of such isotopes can give information on
the fundamental constants at a time when the fossil reactor was
operating. Particularly, the limitations in \eqs{schl} have
appeared because of the resonance
\beq \label{149150}
{}^{149}{\rm Sm} + n \to {}^{150}{\rm Sm} + \gamma\,,
\eeq
which has an energy of only 97.3 meV.
Two isotopes of samarium (147 and 149) have been studied. The halftime of the
isotopes is presented in Table \ref{Tsama}\footnote{Nuclear
data (and particularly
in Tables \ref{Tsama}, \ref{Tmu}  and \ref{Tprop}) have been
taken from Ref. \cite{firestone}, when the reference is not specified.}.
The cross
sections of reaction in \eq{149150} and a similar one for
\beq
{}^{147}{\rm Sm} + n \to {}^{148}{\rm Sm} + \gamma\,,
\eeq
differ by about two orders of magnitude because of the resonance.
Studying the ${}^{147}{\rm Sm}/{}^{149}{\rm Sm}$ ratio one can
deduce a possible variation of the position of the resonance from
when the reactor was operating to the present day. The strength of
the limitations in \eqs{schl} has three reasons:
\begin{itemize}
\item the sensitivity of the abundance of samarium isotopes
to the position of the resonance;
\item the fact that the resonance energy
($\sim 100$ meV) has to be compared with a well of the nuclear
potential ($\sim 50$ Mev) i. e. it is $5\cdot10^8$ times larger
then the energy;
\item large time separation ($\sim 2\cdot 10^9$ yr).
\end{itemize}

\begin{table}[th]
\begin{center}
\begin{tabular}{||c|c|c|c||}
\hline\hline
&&&\\[-1ex]
Isotopes & Halftime & Natural & Neutron separation \\[1ex]
&& abundance & energy $S_n$ [keV]   \\[1ex]
\hline
&&&\\[-1ex]
147 & $1.06\cdot 10^{11}$ yr &  15.0\% & 6342(3)~~~ \\[1ex]
148 & $7\cdot 10^{13}$ yr    &  11.3\% & 8141.5(6) \\[1ex]
149 & $> 2\cdot 10^{15}$ yr  &  13.8\% & 5871.6(9) \\[1ex]
150 & stable                 &  ~7.4\% & 7985.7(7) \\[1ex]
\hline\hline
\end{tabular}
\vspace{5mm} \caption{\em  Properties of some samarium isotopes.
\label{Tsama}}
\end{center}
\end{table}

After publication of Ref. \cite{shlyakhter} the Oklo data have
been re-evaluated by a number of authors
\cite{petrov,irvine,sisterna,damour,fujii}. The evaluations were
concentrated on the samarium abundance. Particularly, it was
pointed out in Ref. \cite{fujii} that for the limitations for the
strong coupling constant it is unlikely to be appropriate to
compare the position of the resonance to the well of the
nucleon-nucleon potential, which is essentially of use only in
the few-body problem. In the case of many-body effect (like the
resonance) it was suggested to consider a neutron separation
energy $S_n$ (see Table \ref{Tsama}) as a characteristic reference
value. The latter is 6--8 Mev and significantly smaller than the
well ($\sim 50$ Mev).

Investigations of other isotopes were not as effective in setting limits.
Some estimations due to europium were presented by Shlyakhter, while
gadolinium was studied in Ref. \cite{fujii}. But the studies did
not yield such strong limitations as the investigations of the
samarium isotopes.

Results for the variation of the fine structure constant from the
Oklo reactor study are collected in Table \ref{Toklo}. The most
recent estimates are \cite{damour}
\begin{equation} \label{eqgeo}
\frac{1}{\alpha}\,\frac{\partial \alpha}{\partial t}=-1.4(54)\cdot 10^{-17}
~{\rm yr}^{-1}\,,
\]
\[
\frac{1}{\alpha_{\sc w}}\,\frac{\partial \alpha_{\sc w}}
{\partial t}<1\cdot 10^{-11}~{\rm yr}^{-1}\,,
\end{equation}
and \cite{fujii}
\begin{equation}
\frac{1}{\alpha}\,\frac{\partial \alpha}{\partial t} < 1.0\cdot 10^{-17}
~{\rm yr}^{-1}\,,
\]
\[
\frac{1}{\alpha_{\sc s}}\,\frac{\partial \alpha_{\sc s}}
{\partial t}<1.3\cdot 10^{-18}~{\rm yr}^{-1}\,.
\end{equation}
Those results cannot be used directly, because of the same reasons
as those for other geochemical data. Particularly, the timing was
based on the abundance of strong absorbers of thermal neutrons and
any influence of the variation of the constants on utilized cross
sections was not investigated. Variation of the fine structure
constant was analyzed only with respect to the static Coulomb
interaction energy of the proton in the nucleus. However, we have
to expect that a small part of the proton and neutron masses is a
result of electromagnetic interactions and hence any variations of
$\alpha$ lead to a shift in these masses and of the kinetic
energy. Indeed only a small part of the kinetic energy is of this
electromagnetic origin. However the entire kinetic energy is much
larger than the static Coulomb interaction and eventually both
electromagnetic contributions can be compatible.

Concerning two different recent results on the fine structure
constant we note that in Ref. \cite{fujii} more recent and
accurate data on the samarium isotopic composition were used,
while in Ref. \cite{damour} the estimation of the temperature was
more secure. We also have to point out that it is unclear how much
the neutron flux during the operation time of the reactor used in
Refs. \cite{damour,fujii} is sensitive to the possible variation
of the constants. We think a proper way is to determine the flux
and a possible variation simultaneously.

\begin{table}[th]
\begin{center}
\begin{tabular}{||c|c||}
\hline\hline
&\\[-1ex]
$\partial ln{\alpha}/\partial t$ & Ref. \\[1ex]
\hline
&\\[-1ex]
$<1\cdot 10^{-17}~{\rm yr}^{-1}$ & \protect{\cite{shlyakhter}} \\[1ex]
$<2\cdot 10^{-18}~{\rm yr}^{-1}$ & \protect{\cite{irvine}} \\[1ex]
$-1.4(54)\cdot 10^{-17}~{\rm yr}^{-1}$ & \protect{\cite{damour}} \\[1ex]
$ < 1.0\cdot 10^{-17}~{\rm yr}^{-1}$ & \protect{\cite{fujii}} \\[1ex]
\hline\hline
\end{tabular}
\vspace{5mm} \caption{\em  Limits for the $\alpha$ variation from
Oklo. \label{Toklo}}
\end{center}
\end{table}

\subsubsection{Nucleosynthesis}

Some estimations due to Big Bang nucleosynthesis have also been
performed \cite{kolb,bergstrom} in a similar way to the
geochemical study. A possible variation of the fine structure
constant is not larger than $(1-2)\cdot 10^{-12}$ yr$^{-1}$. This
cosmological estimate takes an advantage of a large time
separation between the epoch of nucleosynthesis and the present
day which is about the lifetime of the universe, i. e. about
$10^{10}$ yr.

\subsection{Astrophysical data}

\subsubsection{Absorption lines in quasar spectra}

An advantage of astrophysical studies is a possibility of a long
term comparison. A typical astrophysical time associated with
extragalactic sources (quasars) is up to $10^{10}$ yr. Due to the
long reference time the accuracy of spectroscopic measurements
need not be high. It is also possible to look for corrections
(like $m_e/m_p$ terms in H$_2$ spectrum). A key point of any
astrophysical study is a comparison of observed lines with a data
base of lines collected under laboratory conditions in order to
determine a value of the redshift
\beq
\lambda_{\rm obs} = \lambda_0 \big( 1 + z \big)
\eeq
and, hence, a time separation between the epoch of the absorption
and the epoch of the observation $t(z)$. This depends on the
choice of evolution parameters of the universe and can vary by a
factor two for the same $z$. When the lines are identified, one
can try to interpret spectroscopic data in terms of variations of
transition frequencies. It is quite important to determine the
redshift and look for the variation simultaneously. When one does
this separately it is equivalent to an assumption on the stability
of particular transitions. E. g. in Ref. \cite{varshalovich951},
the authors used the redshift from the observation of some astrophysical
data on the hydrogen {\em hfs\/} line, and so they actually
assumed within the evaluation that the $\alpha^2 \mu_p/\mu_{\sc b}
Ry$ is a stable value\footnote{That was pointed out in Ref.
\cite{drinkwater}. \label{fnote}} and this led to some
misinterpretation.

A comparison of frequencies with different scaling behaviours is
described in Sect. \ref{scal}. Particularly, Savedoff
\cite{savedoff} pointed out this application for the comparison of
atomic lines. Thompson first noted that molecular spectra could be
used to examine the variation of the nuclear masses
\cite{thompson}. An analysis of absorption lines of molecular
hydrogen in a quasar spectrum can possibly provide a limit for the
variation of the proton-to-electron mass ratio. Such an
evaluation is based on the Born-Oppenheimer approximation of the
molecular spectrum
\begin{equation}
\nu_{\rm mol} = {\rm Ry} \left[c_e + c_v\left(\frac{m_e}{M}\right)^{1/2}
+ c_r\left(\frac{m_e}{M}\right)+\dots\right]
\,,
\end{equation}
where $c_i$ are dimensionless parameters of order ${\cal O}(1)$.
The dominate term $c_e$ is determined by the electronic structure,
the second is due to the vibrational excitations, while the third
one is associated with the rotational levels. Comparing levels
with the same electronic structure (the same $a$), it is possible
to study two other terms and to limit the variation of the
electron-to-proton mass ratio from the H$_2$ spectrum. Value $M$
is associated with some nuclear mass, particularly in the case of
diatomic molecules it is the reduced mass of two nuclei. It is
important for applications that the coefficients $c_i$ can be
found both theoretically and experimentally. The latter is
possible by studying different isotopes, particularly the reduced
mass $M$ for H$_2$, HD, D$_2$ etc varies enough to allow this (see
e. g. Ref. \cite{varshalovich951}).

An examination of the molecular lines, or a comparison of the
rotational and vibrational transitions with the gross structure
yields a variation of $m_e/m_p$. In contrast to this a comparison
of the rotational terms to the {\em hfs\/} in the hydrogen atom
yields a limit for the $\alpha^2\,g_p$ variation. Analysis of the
molecular lines was performed in Ref.
\cite{varshalovich93,varshalovich96}, while in Refs.
\cite{tubbs,cowie,varshalovich951,drinkwater97,drinkwater} the
authors preferred to compare the rotational and vibrational
transitions with the {\em hfs\/} of the hydrogen atom. The most
accurate results are collected in Table \ref{Tastro}.

A comparison of the gross structure to the fine structure of some
ions presented in Refs.
\cite{bachall,wolfe,levashev,potekhin,varshalovich94,cowie,varshalovich951,varshalovich953,varshalovich96,ivanchik,potekhin98}
gave limitations on the variations of $\alpha^2$. Variations of
the same value can be found after studying the relativistic
corrections. The most recent results were obtained in \cite{webb},
where an evaluation of data for some of Fe$^+$ and Mg$^+$ lines
was performed. The variation of the fine structure constant was
obtained from a study of relativistic corrections, calculations
for which were presented separately in Ref. \cite{dzuba,dzuba99}.

The {\em hfs\/} of atomic hydrogen was examined with respect to
the gross structure in Refs. \cite{wolfe,cowie} and with respect
to the fine structure in Refs. \cite{wolfe,varshalovich951}. The
former of those examinations is for possible variations of
$\alpha^2\,(\mu_p/\mu_{\sc b})$, while the latter is for
$\mu_p/\mu_{\sc b}$. The strongest astrophysical limitations are
summarized in Table \ref{Tastro}, where we give variations for
actually measured values. References to previous, less precise
results can be found in quoted articles and in Ref.
\cite{levashev}.

{
%\small

\begin{table}[th]
\begin{center}
\begin{tabular}{||c|c|c||}
\hline\hline
&&\\[-1ex]
Value & Ref.  &  $\partial \ln(<{\rm value}>)/\partial t$ \\[1ex]
 & &  [yr$^{-1}$] \\[1ex]
\hline
&&\\[-1ex]
$m_e/m_p$ & \protect{\cite{varshalovich93}}& $<3\cdot 10^{-13}$  \\[1ex]
$m_e/m_p$ & \protect{\cite{cowie}} & $-0.8(35)\cdot 10^{-14}$\\[1ex]
$m_e/m_p$ & \protect{\cite{varshalovich953}}& $<2\cdot 10^{-14}$  \\[1ex]
$m_e/m_p$ & \protect{\cite{varshalovich96}}& $9(6)\cdot 10^{-15}$  \\[1ex]
$\alpha^2$ &\protect{\cite{varshalovich94}}
   & $<6\cdot 10^{-14}$\\[1ex]
$\alpha^2$ &\protect{\cite{potekhin}}
   & $<4\cdot 10^{-14}$\\[1ex]
$\alpha^2 $ & \protect{\cite{cowie}} & $1.0(35)\cdot 10^{-14}$\\[1ex]
$\alpha^2$ &\protect{\cite{varshalovich951}}
   & $<3\cdot 10^{-14}$\\[1ex]
$\alpha^2$ &\protect{\cite{varshalovich96}}
   & $<5\cdot 10^{-15}$\\[1ex]
$\alpha^2$ &\protect{\cite{webb,dzuba}}
   & $0.4(51)\cdot 10^{-15}$\\[1ex]
$\alpha^2$ &\protect{\cite{ivanchik}}
   & $<5\cdot 10^{-15}$\\[1ex]
$\alpha^2\,g_p$ &\protect{\cite{varshalovich951}}
   & $<2\cdot 10^{-14}$\\[1ex]
$\alpha^2\,g_p$ &\protect{\cite{drinkwater,drinkwater97}}
   & $<2\cdot 10^{-15}$\\[1ex]
$\alpha^2 \mu_p/\mu_{\sc b}$ & \protect{\cite{wolfe}}
    & $<2\cdot 10^{-14}$\\[1ex]
$\alpha^2 \mu_p/\mu_{\sc b}$ & \protect{\cite{cowie}}
    & $-1.0(13)\cdot 10^{-15}$\\[1ex]
\hline\hline
\end{tabular}
\vspace{5mm}
\caption{\em Possible variations of the constants from astrophysics.
\label{Tastro}}
\end{center}
\end{table}
}

\subsubsection{Background radiation}

An analysis of microwave background radiation data to be obtained
in the near future is expected to give a limitation for a
variation of the fine structure constant of $10^{-12}-10^{-13}$ yr
$^{-1}$ \cite{kaplinghat,hannestad}.

\section{Laboratory search}

The limitations from \eq{eqgeo} and Table \ref{Tastro} are
stronger than possible in any laboratory experiments. However,
to our mind, the most reliable limitations can be achieved only
under laboratory conditions and there are a few very different
ways to determine limits for possible variations of the
fundamental constants.

\subsection{Laboratory nuclear data \label{LabNucl}}

As has been mentioned, Shlyakhter noted that the laboratory study
of some nuclear properties can give reasonable limitations for
variations of the constants \cite{shlyakhter}. He considered some
very low-lying resonances, the energies of which are extremely
small differences of large quantities and those must be sensitive
to small variations of parameters. The positions of some of these
resonances had been known with enough accuracy for about 10 years
and an estimate
\beq \label{nuclabo} \frac{1}{\alpha_{\sc s}}\,\frac{\partial
\alpha_{\sc s}} {\partial t} <4\cdot 10^{-12}~{\rm yr}^{-1}
\eeq
was obtained \cite{shlyakhter}. The difference between this
approach and others was that the most sensitive values were
studied, whereas others investigated easily available data from
geochemistry. Such a test is free of the timing problem, although,
Schlyakhter's analysis has been performed under the assumption of
particle masses stability. Unfortunately, to the best of our
knowledge this idea has not been developed further. Actually the
estimations \cite{shlyakhter} from laboratory data were
competitive with ordinary geochemical data examined that time.
However, that is not ture in the case of the Oklo reactor.

We think that it is necessary to examine the data base of
low-lying resonances. Even, if there is no progress in measurement,
the limitation in \eq{nuclabo} is reduced by a factor of about
3.5. That is a result of adjustment of analysis by Shlyakhter, who
claimed in 1976, that the positions of the resonances had not been
shifted for 10 years.

\subsection{Clock comparison \label{Clocom}}

Another example for a laboratory search would be a comparison of
different clocks looking for any variation during a relatively
short time ($\leq$ 1 year) \cite{godone93,prestage}. Ref.
\cite{godone93} presented  a one-year comparison of hyperfine
structure of Cs to the fine structure of $^{24}$Mg. Another limit
on a possible variation of the ratio of the frequencies from the
same authors \cite{godone95} is $2.6\cdot10^{-13}$ yr$^{-1}$,
though, the data seems to be the same. They neglected relativistic
corrections in their evaluation and gave some interpretation based
on that.

Another recent comparison of the Hg$^+$ clock (based on the {\em
hfs\/} transition) and hydrogen clock for 140 days \cite{prestage}
led to a limit of $1.7\cdot10^{-14}$ yr$^{-1}$. The original
result was presented in terms of the fine structure constant which
was derived from relativistic corrections to the nonrelativistic
formula. The authors of Ref. \cite{prestage} underlined the
importance of relativistic effects. We would like to point out
that the treatment of the Cs {\em hfs\/} in Ref. \cite{godone93}
is just the opposite: while here the relativistic corrections were
neglected \cite{godone93}, the others believe that the corrections
are crucially important \cite{prestage}. The results from
different clock comparisons are collected in Table \ref{Tclock}.
We give here the limits for possible variations of the ratio of
the frequencies and dismiss any original interpretations. We
include also an H--Cs comparison for 1 year at the PTB \cite{ptb}
with a result $5.5\cdot10^{-14}$ yr$^{-1}$ and at U. S. Navy
Observatory \cite{usnavy}, as has been interpreted in Ref.
\cite{prestage}. We would also like to mention a result of Ref.
\cite{turneaure} because the clock was quite different from
others. A 12-days comparison of a Cs clock to a new standard
based on and SCSO (superconductivity-cavity stabilized oscillator
\cite{SCSO}) was performed. The frequency of the standard depends
on its size, which is taken as proportional to the Bohr radius.
This is correct in a nonrelativistic approximation, and it is not
quite clear how to estimate the relativistic corrections.

{
%\small

\begin{table}[th]
\begin{center}
\begin{tabular}{||c|c|c||}
\hline\hline &&\\[-1ex] Transitions & Ref.  &  $\partial \ln(<{\rm
value}>)/\partial t$ \\[1ex]
 & &  [yr$^{-1}$] \\[1ex]
\hline
& &\\[-1ex]
{\em hfs\/} of Cs to SCSO
  &\protect{\cite{turneaure}} & $<1.5\cdot 10^{-12}$ \\[1ex]
{\em fs\/} of $^{24}$Mg to {\em hfs\/} of Cs
  &\protect{\cite{godone93}} & $-2.5(23)\cdot 10^{-13}$ \\[1ex]
{\em hfs\/} of H to {\em hfs\/} of Cs
  &\protect{\cite{ptb}} & $<5 \cdot 10^{-14}$ \\[1ex]
{\em hfs\/} of H to {\em hfs\/} of Cs
&\protect{\cite{usnavy}} & $<5 \cdot 10^{-14}$ \\[1ex]
{\em hfs\/} of Hg$^+$ to {\em hfs\/} of H
  & \protect{\cite{prestage}} & $<2.7 \cdot 10^{-14}$\\[1ex]
\hline\hline
\end{tabular}
\vspace{5mm} \caption{\em Clock comparisons and possible
variations in frequencies of the standards. \label{Tclock}}
\end{center}
\end{table}
}

Ref. \cite{prestage} in the most important one in the table,
because the hydrogen-to-cesium comparisons are taken from there
and because of discussions on the relativistic effects. We discuss
the interpretation of the hyperfine separation in Sect.
\ref{inter}, but here we comment on some statements of Ref.
\cite{prestage}.
\begin{itemize}
\item
The correcting function for the relativistic effects $F_{\rm
rel}(Z\alpha)$ used there was not quite correct. The authors did
not give enough explanations and to briefly discuss their
evaluation we would like to mention a few points:
\begin{itemize}
\item[-] The relativistic correcting function $F_{\rm rel}(Z\alpha)$
was given with some analytic expression, while the nonrelativistic
term was possible to find only within an empirical formulae.
Indeed, that is inconsistent. The function was expected to be
valid for any alkali atoms including hydrogen-like and Li-like
atoms. The result for $F_{\rm rel}(Z\alpha)$ is $n$-independent
and it was claimed to be valid for $S_{1/2}$ levels, while rather
it should be $n$-dependent (see e. g. Ref. \cite{breit}).
Particularly, the result used in Ref. \cite{prestage} is in
disagreement with both $1s$ and $2s$ results for a hydrogen-like
atom with a nuclear charge $Z$ \cite{breit}. In the case of
low-$Z$ the results for a hydrogen-like atom are
\beq \label{nr1}
F_{\rm rel} (Z\alpha) \simeq 1 + \frac{11}{6}\,(Z\alpha)^2+\dots,
\quad {\rm \cite{prestage}}\,,
\eeq
\beq \label{nr2}
F_{1s} (Z\alpha) \simeq 1 + \frac{3}{2}\,(Z\alpha)^2+\dots, \quad {\rm \cite{breit}}\,,
\eeq
and
\beq \label{nr3}
F_{2s} (Z\alpha) \simeq 1 + \frac{17}{8}\,(Z\alpha)^2+\dots, \quad {\rm \cite{breit}}\,.
\eeq
\item[-] Actually, the calculation of the Casimir correction
\beq \label{Fcasimir}
F_{\rm rel}(Z\alpha) =
\frac{3}{\sqrt{1-(Z\alpha)^2}}\,\frac{1}{3-4(Z\alpha)^2}
\eeq
has been performed under the condition that the relativistic
corrections can appear only when the electron is close to the
nucleus and hence they are proportional to a squared value of the
wave function at the origin \cite{casimir}. That is correct for
heavy ($Z\gg 1$) and slightly charged ($z\ll Z$, where $z$ is an
effective charge for a valence electron) alkali atoms. Indeed that
is incorrect for hydrogen. Actually, the corrections for hydrogen
are small, and it is enough to reproduce a correct order of
magnitude.
\item[-] Recalculation in Ref. \cite{dzuba99} gave results for
$\partial \ln(F_{\rm rel}(Z\alpha))/\partial \alpha$,
which are different from the Casimir calculation
\beq \label{Lcasimir}
\frac{\partial \ln\Big(F_{\rm rel}(Z\alpha)\Big)}{\partial\ln \alpha}
 = \frac{(Z\alpha)^2\,\Big(11-12(Z\alpha)^2\Big)}
{\Big(1-(Z\alpha)^2\Big)\,\Big(3-4(Z\alpha)^2\Big)}
\eeq
within about 10\%. The results are 2.30 for Hg$^{+}$ and 0.83 for
Cs \cite{dzuba99} instead of 2.2 and 0.74 \cite{prestage}.
Actually the mercury ion is not an alkali one, but since all
subshells are closed the Casimir approximation has to work and
that has been confirmed by the many-body calculation
\cite{dzuba99}.
\end{itemize}
\item The authors assumed that there are no corrections to the
nuclear $g$-factor which depend on the strong coupling constant.
Actually that means that the magnetic moment of any nucleus is to
be understood in terms of a pure kinematic description (spin and
orbital contributions) with high accuracy. That is definitely not
the case (see Sect. \ref{inter} for detail). The corrections do
not grow with increase of the nuclear charge $Z$, but nevertheless
they are large enough for a number of values of $Z$ in a broad
range. Particularly, for tritium ($Z=1$) such effects shift a
value of the nuclear magnetic moment by 7\% (cf. \eq{Fcasimir}
with $Z=26$). Even in the case of a pure kinematic model, it is
incorrect to neglect variations of $g_p$ and $g_n$, which
contribute differently to magnetic moments of hydrogen, rubidium,
cesium and mercury (see Sect. \ref{inter} for detail).
\item The hydrogen maser was tried intensively
as a candidate for the primary frequency standard about 30 years
ago. A crucial problem was low long-time stability and it has not
been improved up-to-now. That means that the maser frequency can
disagree with the transition frequency and varies with the time
because of different effects, particularly, a wall-shift.
Comparison of anything with the hydrogen maser itself makes no
sense. Actually, the authors of Refs. \cite{ptb,usnavy} make no
statement on a possible variation of any transition frequency, and
the interpretation in Table \ref{Tclock} is from Ref.
\cite{prestage}.
\end{itemize}

The best limits for the annual variations of the constants from
the clock (see Table \ref {Tclock}) are on the level of a few
units in $10^{14}$ but it is not quite clear if any direct
interpretation of such a comparison is actually possible. A clock
is a device designed to maintain some frequency in the most stable
way. An equality of the maintained frequency to any atomic
transition frequency is not necessary, and, actually, it is not
quite clear, if any clock frequency agrees with the transition one
within an accuracy on the level of its reproducibility. We expect
that, if it were to agree it would be no problem to have better
limitations by just searching for a longer time.

In particular, we expect that variations of a frequency of any
maser standards should also be determined by the cavity size (cf.
SCSO standard \cite{turneaure}). The frequency of the hydrogen
{\em hfs\/} transition is not itself important in some sense for
the hydrogen maser. When the {\em hfs\/} frequency and the size of
the resonator are inconsistent the standard cannot work, and when
they are consistent (within the line width) everything is
determined by the cavity. That is indeed a reason why there are a
lot of possibilities of drifts for the frequency from the maser
standard. Such a maser, without any tuning of the cavity size is
called a {\em passive maser\/}. On the other hand in the case of
{\em active maser\/} there is an adjustment of the cavity size to
the hydrogen {\em hfs\/} studying the efficiency of the maser. In
this case the dependence on the variation of the constants is more
complicated.

It is known that hydrogen masers can have high short-time
stability. But there is no {\em a priori\/} statement applicable
to any particular hydrogen maser, after a rather preliminary study
of these. The study assumes some comparisons with either another
known standard, or a wide representative ensemble of masers. In
both cases, any later comparison with that maser assumes that
is consistent with some other standard and that the stability
property (e. g. short-term stability) does not vary with time.
This is an indirect comparison with something else via the maser.
In the case of a 140 day comparison the maser stability and
agreement between the maser frequency and the hydrogen {\em hfs\/}
are questionable and the stability may only be a result of
preliminary study of the maser frequency with respect to some
other standard.

We think that the clock comparison can give reliable limits only
in two cases:
\begin{itemize}
\item The clock frequency is expressed in terms of the transition
frequency. But that means, that with the clock comparison one can
measure the transition frequency as well. We consider some of
possibility for search with the precise frequency measurements
below (see Sect. \ref{Prec}).
\item There are a number of different standards with the same transition
(like in the case of Cs). They should first be compared with each
others and then we can estimate possible deviations from some
effective frequency that depends on the transition rather than on
the clock. This is not as secure as a direct determination of the
transition frequency, but it is more or less reliable.
\end{itemize}
In the case where  some particular transition is applied in only a
single clock, it is absolutely unclear, which drifts or
fluctuations are properties of the clock and which are properties
of the transition.

\subsection{Precise frequency measurements \label{Prec}}

Precision spectroscopy provides us with other ways to search for
the variation of the fundamental constants. The most
straightforward method is to obtain a high accuracy and to compare
two results (let us say, one taken a year after the other). It is
also possible to make a comparison of results obtained now with
some relatively old ones for the frequency of atomic or molecular
transitions measured better than $10^{-11}$. Tables of the most
accurately measured values of any transition frequencies are
presented below. Table \ref{Trf} contains the best radiofrequency
results, while Table \ref{Topt} is for the optical transitions.
The tables contain the results obtained from 1967 to the
present with a fractional uncertainty below $10^{-11}$. One can
see that some older results are competitive with the newer ones.
In the tables we give a possible limit of variation of the
frequency if the new experimental value is to be obtained in the
year 2000 and with some higher accuracy. If the precision is about
the same, one should also take into consideration an uncertainty
in the newer measurement. So, the final limit has to be larger
than that given in the tables by a factor between $\sqrt{2}$ (if
the uncertainties are independent) and 2 (when they are strongly
correlated). In our evaluation we consider the date of publication
as the date of the measurement, whereas they are slightly
different and some shifts may arise from this.

{
%\small

\begin{table}[th]
\begin{center}
\begin{tabular}{||c|c|c|c|c||}
\hline\hline
&&&&\\[-1ex]
Atom & Frequency ($\nu$) &
Ref.  & Fractional & $\partial \ln\nu/\partial t$ \\[1ex]
& [kHz]& & uncertainty ($\delta$) &  [yr$^{-1}$] \\[1ex]
\hline
&&&&\\[-1ex]
H  & ~1 420 405.751 766 7(9)~&
  \protect{\cite{ramsey}}, $\sim$1970 &
     $6.4 \cdot 10^{-13}$ & $2.2\cdot 10^{-14}$\\[1ex]
D  & ~~~327 384.352 521 5(17) &
  \protect{\cite{wineland}}, 1972 &
     $5.2\cdot 10^{-12}$ & $1.8\cdot 10^{-13}$\\[1ex]
T & ~1 516 701.470 773(8)~~~ &
  \protect{\cite{mathur}}, 1967 &
     $5.3\cdot 10^{-12}$ & $1.6 \cdot 10^{-13}$\\[1ex]
$^9$Be$^+$ & ~1 250 017.678 096(8)~~~ &
  \protect{\cite{bollinger}}, 1983 &
     $6.4\cdot 10^{-12}$ & $3.8\cdot 10^{-12}$\\[1ex]
$^{87}$Rb & ~6 834 682.610 904 29(9) &
  \protect{\cite{bize}}, 1999 &
     $1.3\cdot 10^{-14}$ & $1.3\cdot 10^{-14}$\\[1ex]
$^{133}$Ba$^+$ & 9 925 453.554 59(10)~~&
 \protect{\cite{knab}}, 1987 &
     $10\cdot 10^{-12}$ & $7.7\cdot 10^{-13}$\\[1ex]
$^{171}$Yb$^+$ & 12 642 812.118 471(9)~~~&
  \protect{\cite{tamm}}, 1993 &
     $7.5 \cdot 10^{-13}$ & $1.1\cdot 10^{-13}$\\[1ex]
       & 12 642 812.118 466(2)~~~&
  \protect{\cite{fisk}}, 1997 &
     $2.5 \cdot 10^{-13}$ & $8.3\cdot 10^{-14}$\\[1ex]
$^{173}$Yb$^+$ & 10 491 720.239 55(9)~~~~&
  \protect{\cite{munch}}, 1987 &
     $8.6 \cdot 10^{-12}$ & $6.6\cdot 10^{-13}$\\[1ex]
$^{199}$Hg$^+$& 40 507 347.996 841 6(4)~&
  \protect{\cite{berkland}}, 1998 &
$1.1\cdot 10^{-14}$ & $0.53\cdot 10^{-14}$\\[1ex]
\hline\hline
\end{tabular}
\vspace{5mm} \caption{\em The most precise measurements of the
ground state hyperfine structure interval and a possible level of
limit for the variation of their frequencies. Values for $\partial
\ln\nu/\partial t$ have been calculated under the condition that
the frequency will be re-measured in the year 2000 with a better
accuracy. The variation of the frequencies is assumed with respect
to the cesium {\em hfs\/}. \label{Trf}}
\end{center}
\end{table}
}

{
%\small

\begin{table}[th]
\begin{center}
\begin{tabular}{||c|c|c|c|c||}
\hline\hline
&&&&\\[-1ex]
Transition & Frequency ($\nu$) &
Ref.  & Relative & $\partial \ln\nu/\partial t$ \\[1ex]
& [kHz]& & uncertainty &  [yr$^{-1}$] \\[1ex]
\hline
&&&&\\[-1ex]
H, ${\rm 1s}-{\rm 2s}$ & 2 466 061 413 187.34(84)~~&
   \protect{\cite{udem}}, 1997&
     $3.4\cdot 10^{-13}$ & $1.1\cdot 10^{-13}$\\[1ex]
H, ${\rm 2s}-{\rm 12d}$ & 799 191 727.402 8(67)~~&
   \protect{\cite{schwob}}, 1999&
     $8.4\cdot 10^{-12}$ & $8.4\cdot 10^{-12}$ \\[1ex]
D, ${\rm 2s}-{\rm 12d}$ & 799 409 184.967 6(65)~~&
   \protect{\cite{schwob}}, 1999&
     $8.1\cdot 10^{-12}$ & $8.1\cdot 10^{-12}$ \\[1ex]
$^{40}$Ca, $^3{\rm P}_1-{}^1{\rm S}_0$ & ~~455 986 240 493.95(43)~~ &
  \protect{\cite{schnatz}}, 1996 &
     $9.4\cdot 10^{-13}$ & $2.4\cdot 10^{-13}$\\[1ex]
  & ~~455 986 240 494.13(10)~~ &
  \protect{\cite{riehle}}, 1999 &
     $2.5\cdot 10^{-13}$ & $2.5\cdot 10^{-13}$\\[1ex]
$^{88}$Sr$^+$, ${\rm 5S}-{\rm 4D}$ & ~~444 779 044 095.4(2)~~~~ &
  \protect{\cite{bernard}}, 1999 &
     $4.5\cdot 10^{-13}$ & $4.5\cdot 10^{-13}$\\[1ex]
CH$_4$, E-line & ~~~88 373 149 028.53(20)&
  \protect{\cite{ering}}, 1998 &
     $2.3\cdot 10^{-12}$ & $1.1\cdot 10^{-12}$\\[1ex]
\hline\hline
\end{tabular}
\vspace{5mm} \caption{\em The most accurate optical measurements
and a possible level of limit for the variation of the frequency.
Values for $\partial \ln\nu/\partial t$ have been found under the
condition that a measurement of the frequency will be repeated in
the year 2000 with a higher precision. \label{Topt}}
\end{center}
\end{table}
}

The hydrogen {\em hfs\/} is presented in Table \ref{Trf} with a
value from review \cite{ramsey}. We discuss the original results
in Sect. \ref{Hhfs}. The tables mainly indicates some
opportunities for experiments in the near future. Except for the
hydrogen {\em hfs} only one value in Table \ref{Trf} has been
accurately and independently measured twice (namely, the {\em
hfs\/} intervals in the $^{171}$Yb$^+$ atom). A comparison of two
$^{171}$Yb$^+$ measurements (\cite{tamm} and \cite{fisk}) gives
the variation of the frequencies
\beq
\frac{\partial }{\partial t}
\ln\left(\frac{
\nu_{\sc hfs}({\rm ^{171}Yb^+})}{\nu_{\sc hfs}({\rm Cs})}\right) \simeq
-1(2)\cdot 10^{-13}\, {\rm yr}^{-1}\,,
\eeq
if we suppose that the time separation is 4 years.

It is important to note, that even a single laboratory study can
give a relatively secure result. In some experiments several traps
were used and so they contain an independent measurement in part.
Since most of the recent precision investigations have the
building of a new standard as a target, some long-term monitoring
of the measured frequency was often performed. Unfortunately, the
published data are rather incomplete, but we expect that some
limits on a level between $10^{-12}$ and $10^{-13}$ yr$^{-1}$ will
be available after complete publications of studies giving most of
the recent results in Tables \ref{Trf} and \ref{Topt}.

The most precise comparison with one of the results that is
already known for some time can be performed for the hydrogen
ground state hyperfine structure interval. The possibility to
reach a good result has almost been missed. However, in the case
of new experiments, like for rubidium or mercury, it may easily be
a shift of 1--2 sigma afterwards. On the other hand, some new
results are going to be presented soon: for the Rb {\em hfs\/}
(better than 10$^{-14}$) and for the $1s-2s$ transition in
hydrogen (a few units in 10$^{14}$). That means that in the case
of any delay the measurement could be not compatible.

It is also important to underline that so-called
variation-of-constants experiments check different possibilities
associated with drifts of primary and secondary standards. A
one-year comparison, which can usually be realized in a
laboratory, is not the same as a kind of ``world wide'' comparison
over years. The hydrogen {\em hfs\/} interval is a value which can
be measured in a number of different laboratories now and which
was in the past also studied in a few different places.

Two radio-frequency measurements (namely for tritium and barium)
were performed for unstable isotopes ($T_{1/2}({\rm T})=12.3$ yr
and $T_{1/2}({\rm Ba})=10.5$ yr) and this indicates that the
search for appropriate transitions should not be limited to stable
isotopes only. We do not mention the radioactivity of rubidium-87
which has a halftime of $4.8\cdot 10^{10}$ yr, comparable with the
age of the universe.

We also have to mention an experiment with the ground state
hyperfine structure of $^9$Be$^+$ in a strong magnetic field. A
splitting between ($M_I=-3/2,\,M_J=+1/2)$ and ($-1/2,\,+1/2)$ was
determined \cite{bollinger85}
\beq \label{beplus}
\nu = 303\,016.377\,265\,070(57)~{\rm kHz}\,.
\eeq
The measurement was performed at a field of about $0.8194\,${\sl
T} and the splitting was found at its magnetic-field-independent
point. The fractional uncertainty is $1.9\cdot10^{-13}$ and it was
believed \cite{bollinger91} that this may be reduced to about
$1\cdot10^{-13}$. A further measurement of the splitting in the
year 2000 is to give a limit of the variation of the Be-frequency
with respect to the cesium standard on level of
$1.3\cdot10^{-14}$.

\begin{figure}[h]
\epsfxsize=12cm \centerline{\epsfbox{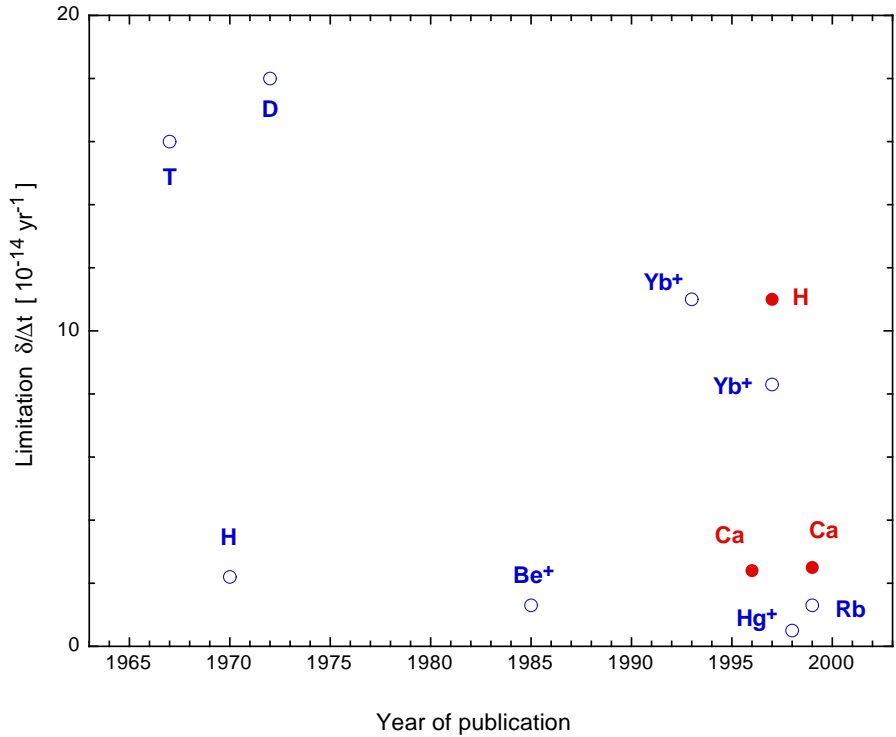}}
\caption{\label{Flim} \em Strongest possible limits on the
variation of the frequencies available in 2000. The limit is
defined as a fractional uncertainty of the frequency divided by
the time from the publication to 2000.}
\end{figure}

In Tables \ref{Trf} and \ref{Topt} and \eq{beplus} in Fig.
\ref{Flim} we collect all limits of the variations of frequency
available in 2000 in the case of a repetition of the measurements.

\section{Some new options for precise comparison of frequencies}

\subsection{Hyperfine structure \label{Shfs}}

For a while the hyperfine splitting of atomic levels has been a
quantity available for the most precise measurements. Comparison
of {\em hfs\/} in different atoms can give us precise information
on the variation of the nuclear magnetic moment rather than on the
fine structure constant.

\begin{itemize}
\item We start with the hydrogen {\em hfs\/} project.
The hyperfine structure interval in the ground state of the
hydrogen atom was frequently measured (see Sects. \ref{Histo} and
\ref{Hhfs}). For a preliminary estimation we accept a value
\begin{equation} \label{average}
\nu_{\sc hfs}({\rm H})= 1420\, 405.751\, 766\, 7(9)\; {\rm kHz}\,,
\end{equation}
which used to be presented in reviews (see e. g. Ref.
\cite{ramsey}) as a final result for the {\em hfs\/}
separation. The fractional uncertainty is about 6 parts in
$10^{14}$ and on being divided by 30 years that gives $2\cdot
10^{-14}$ yr$^{-1}$. If the accuracy is now the same this should
rather be multiplied by $\sqrt{2}$.
\item Let us consider briefly a possible variation of the deuterium hyperfine
separation, which was measured in 1972 \cite{wineland} with an
uncertainty of $5.2\cdot 10^{-12}$. The relative accuracy is much
worse than for H (cf. \eq{average}). However, the magnetic moment
of a deutron (see Table \ref{Tprop}) includes a large cancellation
\beq \label{mud}
\mu({\rm D}) \simeq \mu_p+\mu_n = |\mu_p| - |\mu_n|
\eeq
between the proton ($\mu_p=2.793\,\mu_{\sc n}$) and neutron
($\mu_n=-1.913\, \mu_{\sc n}$) contributions and this {\em hfs\/}
value might be very sensitive to the variation of effective
parameters of the strong interactions.
\item The {\em hfs\/} of the ground state in the $^{171}$Yb$^+$ ion
can also provide a limit for the variation per year on a level of
a few units in $10^{14}$. The most precise measurements are
presented in Fig. \ref{Fyb}, where the open circles are for
preliminary results \cite{bauch,sellars}, while the full ones are
for final values \cite{tamm,fisk}.

\begin{figure}[h]
\epsfxsize=8 cm \centerline{\epsfbox{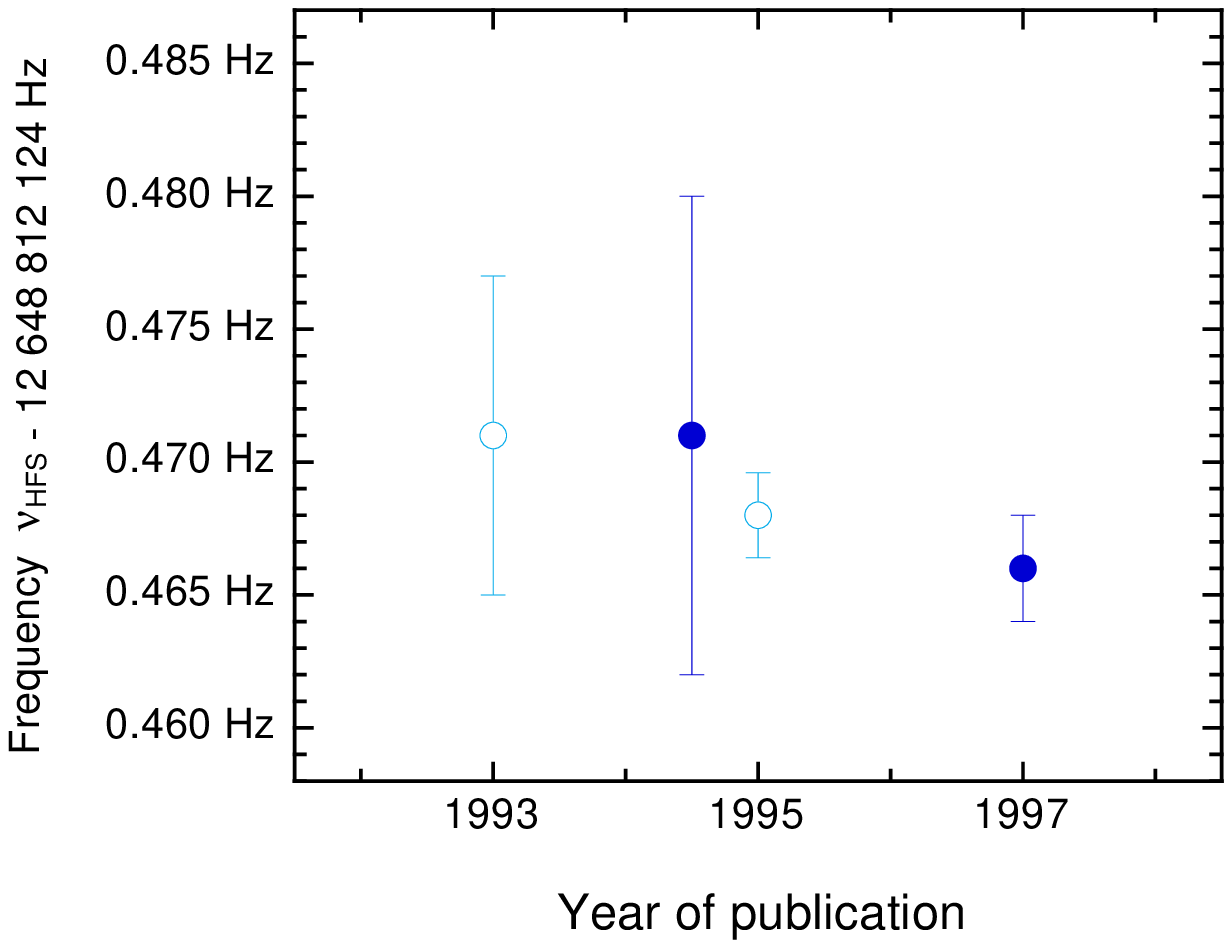}} \caption{\label{Fyb}
\em Precise study of the {\em hfs\/} interval in the ground state
of $^{171}Yb$.}
\end{figure}

The disadvantages (in comparison to the hydrogen case) are: a less
strong limit for the variation with less reliability (the result
was reached with a high accuracy in two laboratories, but the
precision was different by a factor 3).
\item Generally, study of nuclei with small magnetic moments are expected to
lead to sensitive tests for possible variations of the proton or
neutron magnetic moment. We give a list of stable nuclei with
small magnetic moments in Table \ref{Tmu}. The tungsten
$^{183}_{~74}$W has a halftime of $1.1\cdot 10^{17}$ yr and we
include it in the table.

\begin{table}[th]
\begin{center}
\begin{tabular}{||c|c|c|c|c||}
\hline\hline
&&&&\\[-1ex]
      &         & Natural   & Nuclear  & Magnetic \\[1ex]
~$Z$~ & Isotope & abundance & spin and &  moment \\[1ex]
      &         &           & parity   & [$\mu_{\sc n}$]\\[1ex]
\hline
&&&&\\[-1ex]
~7 & $^{15}$N   & 0.37 \%  & $1/2^-$ & -0.283\,188\,4(5) \\[1ex]
19 & $^{41}$K   & 6.7 \%  & $3/2^+$ & 0.214\,870\,1(2)   \\[1ex]
26 & $^{57}$Fe  & 2.2 \%   & $1/2^-$ & 0.090\,44(7)   \\[1ex]
39 & $^{89}$Y   & 100 \%   & $1/2^-$ & -0.137\,415\,4(4) \\[1ex]
45 & $^{103}$Rh & 100 \%   & $1/2^-$ & -0.088\,40(2)  \\[1ex]
47 & $^{107}$Ag & 52 \% & $1/2^-$ & -0.113\,57(2) \\[1ex]
   & $^{109}$Ag & 48 \% & $1/2^-$ & -0.130\,56(2) \\[1ex]
64 & $^{155}$Gd & 15 \%  & $3/2^-$ & -0.259\,1(5)   \\[1ex]
   & $^{157}$Gd & 16 \%  & $3/2^-$ & -0.339\,8(7)   \\[1ex]
69 & $^{169}$Tm & 100 \%   & $1/2^+$ & -0.231\,6(15)   \\[1ex]
74 & $^{183}$W  & 14 \%  & $1/2^-$ &  0.117\,784\,76(9)  \\[1ex]
76 & $^{187}$Os  & 1.6 \%  & $1/2^-$ &  0.064\,651\,89(6)  \\[1ex]
77 & $^{191}$Ir & 37  \%   & $3/2^+$ & ~0.150\,7(6)   \\[1ex]
   & $^{193}$Ir & 63  \%   & $3/2^+$ & ~0.163\,7(6)   \\[1ex]
79 & $^{197}$Au & 100 \%   & $3/2^+$ & 0.145\,746(9)   \\[1ex]
\hline
\hline
\end{tabular}
%\vspace{5mm}
\caption{\label{Tmu} \em Isotopes with small nuclear spin.}
\end{center}
\end{table}

Up to now, there is no way to measure the nuclear magnetic moment
with an accuracy on the level $10^{11}$ or better.  For nuclei
with spin 1/2, our proposal is to measure the {\em hfs\/} of a
neutral atom or ion and to search for a variation of that value.
In the case of spin 3/2, a value of the {\em hfs\/} interval is 
essentially affected by the nuclear quadrupole moment. An
exception is $^{41}$K, where the quadrupole term is also small and
it is worthwhile studying the {\em hfs\/}. Fortunately, the stable
nuclei with the smallest magnetic moments (namely, $^{57}$Fe,
$^{103}$Rh and $^{187}$Os) have spin 1/2 and we hope that the
accurate study of the {\em hfs\/} of the nuclei with a small
magnetic moment is possible.

The nuclear magnetic moment of radionuclides can be even smaller,
for example, for $^{198}_{~81}$Tl ($|\mu| < 10^{-3}\,\mu_{\sc n}$,
$I=2^-$, $T_{1/2}=5.3(5)$ h), $^{153}_{~62}$Sm ($\mu = -
0.022\,\mu_{\sc n}$, $I=3/2^-$, $T_{1/2}=46$ h) and
$^{192}_{~79}$Au ($\mu = - 0.009(2)\,\mu_{\sc n}$, $I=1^-$,
$T_{1/2}=4.9$ h). Our proposal for such isotopes is to measure a
value of the shielded magnetic moment of the nucleus,
investigating ions with coupled electrons only (Hg-, Os-, W-,
Hf-, Ba-, Xe-like etc). Particularly, Hg-, Ba- and Xe-like
ions have complete subshells and this is an advantage for the
study of $^{198}$Tl. We hope that a method developed in Ref.
\cite{gfactor} to study a bound electron $g$-factor in H-like
ions can be applied here. By achieving an uncertainty of about
$10^{-9}$ for the magnetic moment, the limit for the variation of
$g_p$ is expected to be on the level of $3\cdot 10^{-13}$
yr$^{-1}$. We should mention that not all magnetic moments of
radionuclides with a halftime longer than 10 days are known
\cite{firestone} and it might happen, that some of them are even
smaller than $10^{-3}\,\mu_{\sc n}$.
\end{itemize}

\subsection{Fine structure}

When the fine structure (proportional to $\alpha^2Ry$) is
determined, it can be compared with the gross structure
(proportional to $Ry$) and thus yielding a direct limit for a
variation of the fine structure constant $\alpha$. The gross
structure can be taken from measurements with neutral hydrogen and
calcium atoms, and with strontium and indium ions and, maybe, in
the future with other atoms.

\begin{itemize}
\item To date there are no competitive results on the {\em fs\/}.
The best result for the atomic fine structure has been reached for
the Ba$^+$ ion \cite{madej}\footnote{In Refs.
\cite{godone93,godone95} a clock, based on fine structure in Mg,
was compared with a cesium clock. However, the authors gave no
result of the {\em fs\/} transition frequency. As I was informed
by A. Godone it was expected that the corrections to the  {\em
fs\/} were well understood at least on level of $10^{-12}$.}
\beq \label{bafs}
\nu_{\rm fs}(^{138}{\rm Ba^+},\; 5d{}^2D_{3/2}-5d{}^2D_{5/2})
= 24\,012\,048\,317.17(44)\,{\rm kHz}\,.
\eeq
Note that this is for the {\em  fs\/}  of excited states. We think
that some higher accuracy can be achieved by studying the fine
structure associated with the ground state. If the subshell with
valence $p$-electrons (or $d$) is open, the lowest excited states
are due to the fine structure. The frequency can lie in
radio-frequency range and the lines are very narrow. That is
because of two reason: the $E1$ transition is not allowed for
$P-P$ transitions and the decay rate is proportional to some power
of the low transition frequency. Thus the lowest levels are split
due to relativistic effects only and they can be measured
accurately. This is another way of measuring the fine structure
precisely. In combination with the gross structure, one can reach
a limit for $\alpha$. The interpretation of such a comparison is
simpler than in the case of {\em hfs\/}. A similar way is to study
the relativistic correction for the gross structure
is of use for 
astrophysical data \cite{savedoff}. In Table
\ref{Tatom}\footnote{Atomic data (and particularly that in Tables
\ref{Tatom}, \ref{Tion} and \ref{Tag}) have been taken from Ref.
\cite{radzig}, unless otherwise specified.} we give a list of the
{\em rf\/} transition of the low-lying fine structure in some
neutral atoms, while those for different ions are collected in
Table \ref{Tion}.

\begin{table}[th]
\begin{center}
\begin{tabular}{||c|c|c|c|c|c||}
\hline\hline
&&&&&\\[-1ex]
~$Z$~ & Atom & Level & Energy & Lifetime & Nuclear spin \\[1ex]
\hline
&&&&&\\[-1ex]
~5  & B(2$^2$P$_{1/2}^0$) & 2$^2$P$_{3/2}^0$ & 0.457 THz
       & $3.2\cdot 10^7$ s & 3 ($^{10}$B), 3/2 ($^{11}$B) \\[1ex]
~6  & C(2$^3$P$_0$)       & 2$^3$P$_1$       & 0.492 THz
       & $1.3\cdot 10^7$ s & 0 ($^{12}$C), 1/2 ($^{13}$C) \\[1ex]
   &                     & 2$^3$P$_2$       & 1.30~ THz
       & $3.7\cdot 10^6$ s &                              \\[1ex]
14 & Si(3$^3$P$_0$)      & 3$^3$P$_1$       & 2.31~ THz
       & $1\cdot 10^5$ s   & 0 ($^{28}$Si, $^{30}$Si), 1/2 ($^{29}$Si) \\[1ex]
\hline
\hline
\end{tabular}
%\vspace{5mm}
\caption{\label{Tatom} \em Low-lying\/ {\rm rf} fine structure of neutral
atoms.}
\end{center}
\end{table}

\begin{table}[th]
\begin{center}
\begin{tabular}{||c|c|c|c|c||}
\hline\hline
&&&&\\[-1ex]
~$Z$~ & Atom & Level & Energy  & Nuclear spin \\[1ex]
\hline
&&&&\\[-1ex]
~6  & C$^{+}$(2${}^2$P$_{1/2}^0$)     & 2${}^2$P$_{3/2}^0$        & ~1.90 THz
     & 0 ($^{12}$C), 1/2 ($^{13}$C) \\[1ex]
~7  & N$^{+}$(2${}^3$P$_0$)           & 2${}^3$P$_1$              & ~1.46 THz
     & 1 ($^{14}$N), 1/2 ($^{15}$N) \\[1ex]
   &                                & 2${}^3$P$_2$              & ~3.92 THz
     &                                \\[1ex]
21 & Sc$^{+}$(3d4s$-{}^3$D$_1$)       & 3d4s$-{}^3$D$_2$            & ~2.03 THz
     & 7/2 ($^{21}$Sc) \\[1ex]
   &                                & 3d4s$-{}^3$D$_3$           & ~5.33 THz
     &                               \\[1ex]
22 & Ti$^{+}$(3d$^2$4s$-{}^4$F$_{3/2}$) & 3$d^2(^3$F)4s$-{}^4$F$_{5/2}$ & ~2.82 THz
     & 0 ($^{46}$Ti, $^{48}$Ti, $^{50}$Ti),  \\[1ex]
   &                                & 3d$^2$4s$-{}^4$F$_{7/2}$      & ~6.77 THz
     & 5/2 ($^{47}$Ti), 7/2 ($^{49}$Ti)      \\[1ex]
23 & V$^{+}$(3d$^4-{}^5$D$_0$)        & 3d$^4-{}^5$D$_1$           & ~1.08 THz
     & 6 ($^{50}$V), 7/2 ($^{51}$V) \\[1ex]
   &                                & 3d$^4-{}^5$D$_2$           & ~3.20 THz
     & \\[1ex]
   &                                & 3d$^4-{}^5$D$_3$           & ~6.26 THz
     & \\[1ex]
   &                                & 3d$^4-{}^5$D$_4$           & 10.17 THz
     & \\[1ex]
\hline
\hline
\end{tabular}
%\vspace{5mm}
\caption{\label{Tion} \em Low-lying\/ {\rm rf} fine structure of
single-charged ions.}
\end{center}
\end{table}

Unfortunately, there are no systems in the tables which can be
easily studied. In the case of neutral atoms in Table \ref{Tatom},
laser cooling is hard to apply because of the metastable fine
structure levels. Finding levels which are insensitive to the
magnetic field is a problem for ions. These are (2${}^3$P$_0$) in
N$^{+}$ and (3d$^4-{}^5$D$_0$) in V$^{+}$. Finding proper means of
detection of the {\em fs\/} transition can be another problem for
a few-ions trap experiments.
\item It is also possible to find an optical or infrared transition for the
low-lying fine structure. Let us mention a transition
\beq
\nu_{\sc fs}({\rm Pb}, 6{}^3P_0 - 6{}^3P_2) \simeq 1.32~{\rm eV}
\eeq
in neutral lead, which can be studied by means of two-photon
Doppler-free spectroscopy (the wave length of each photon is 1.88
$\mu$m). The excited level $^3P_2$ lives for 2.6 {\em s} and it is
narrow enough to reach an accurate result. It has to be mentioned
that in the case of neutral lead, calculations using $jj$ coupling
are competitive with those using $LS$ one. Actually, a clear separation
of non-relativistic and relativistic physics is only possible for
$LS$ coupling. $LS$ coupling means that one must first find a
non-relativistic energy level with $n^*(L)$ and next to take into
account the (relativistic) spin effects. For $jj$ coupling the
(relativistic) spin effects for individual electrons are more
important than a (non-relativistic) interaction of their orbital
momenta. We expect large relativistic corrections to the fine
structure.

\begin{figure}[h]
\epsfxsize=16cm \centerline{\epsfbox{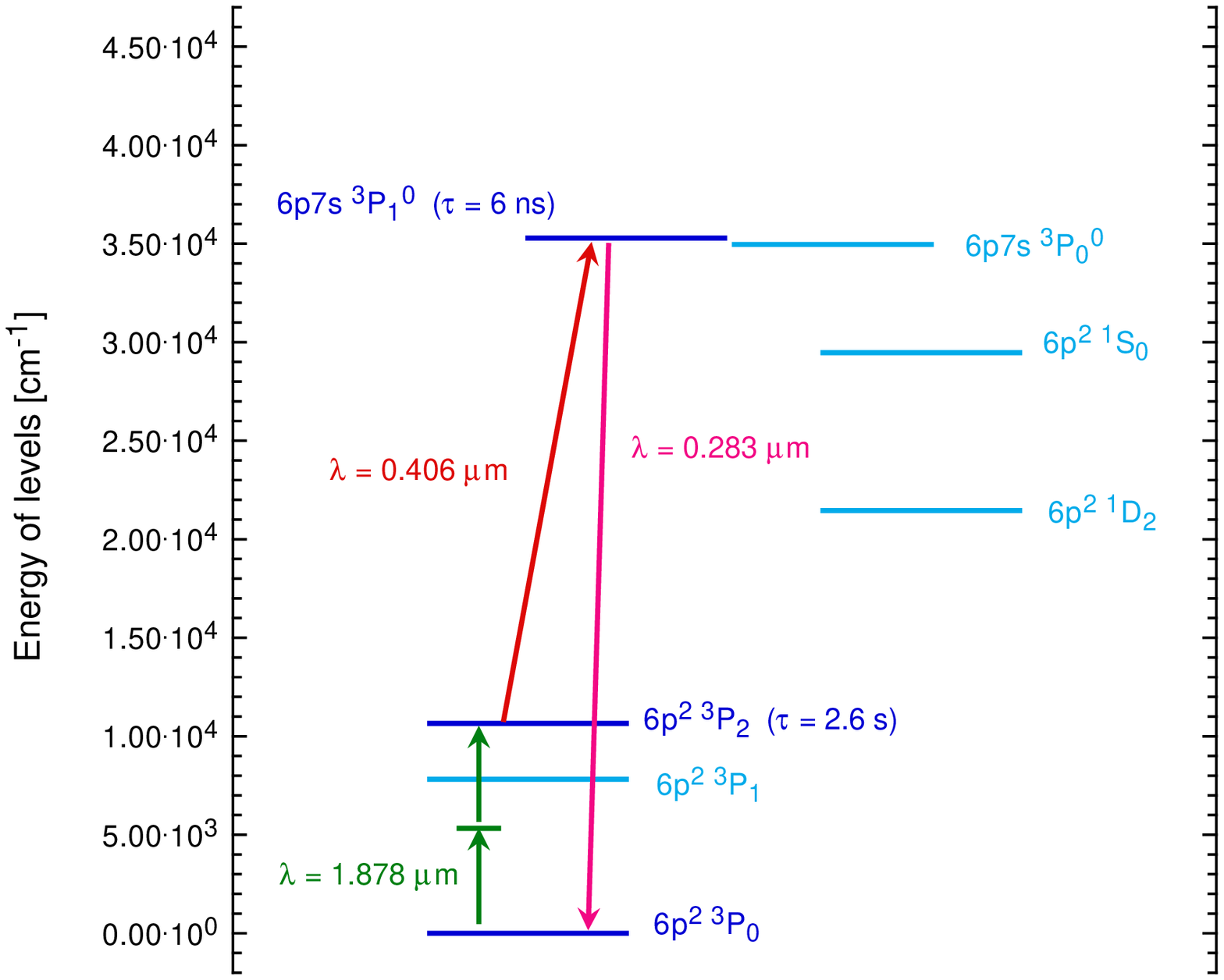}} \caption{\label{Fpb}
\em The energy levels and a scheme of the experiment on the fine
structure of neutral lead.}
\end{figure}

One possible experiment is presented in Fig. \ref{Fpb}. The ground
state ($6p^2{\,}^{3\!}P_0$) of one of the spinless isotopes of
lead ($^{204}$Pb, natural abundance 1.4\%; $^{206}$Pb, 24\%; or
$^{208}$Pb, 52\%) is excited by two photons ($\lambda=1.878\,\mu
m$) to a metastable $^3P_2$ level with a lifetime of 2.6 s. The
line is narrow because of the metastability and of Doppler-free
excitation. Detection of the $^3P_2$ level can be done using an
additional one-photon excitation ($\lambda=0.283\,\mu m$) to a
$6p7s{}^3P_0$ state and measuring the fluorenscence
($\lambda=0.406\,\mu m$).
\item Alkali atoms have simple spectra and that is an advantage for
both experiment \cite{godone93,schnatz,bernard,madej,zanthier} and
theory \cite{casimir,dzuba99}. Measurement of the fine structure
of such a system as a test for the variations of $\alpha$ was
proposed by Jungmann \cite{jungmann} (cf. Ref. \cite{savedoff})
particularly for Ca$^+$ and Sr$^+$ ions. Similar measurement can
be performed for In$^+$. All these atoms are now a subject of some
investigations as a part of efforts to design new optical standards.
Let us discuss
shortly the indium case. An accurate result for the fine structure
may be obtained by considering the $5s5p-{}^3P_J$ levels in the
$^{115}$In$^+$ ion. The {\em fs\/} interval of excited levels
cannot usually be measured precisely. For an indium ion it may
however be determined as the difference between two gross
transitions ($5s^2\!-\!{}^1S_0-5s5p\!-\!{}^3P_J$)ith different $J$.
Indeed, since
the {\em fs\/} interval is about 1\% of the transition frequency
for the gross structure, one can expect that the fractional
accuracy is not very high. The advantage is that in the case of
very accurate measurements of $5s^2\!-\!{}^1S_0-5s5p\!-\!{}^3P_0$
for different $P_J$-states, it may be possible to go beyond the
accuracy of standards. As an example, let us recall the results on
the $1s-2s$ transition in the hydrogen atom \cite{udem} and on the
hydrogen-deuterium isotopic shift of the $1s-2s$ frequency
\cite{huber}. By comparing the two frequencies, it was possible to
detect a drift of the standard used and to reduce the absolute
uncertainty to 150 Hz for the isotopic shift, while for hydrogen
this was 850 Hz. The absolute measurement in the indium ion
($5S-5P_0$ transition) now gives \cite{zanthier} 1 267 402 452
914(41) kHz (uncertainty is $3.2\cdot 10^{-11}$) and the result is
soon to be improved.
\item Another approach to compare the gross and fine structure may
possibly be realized in an atomic system, where the levels with
different $n$ and $l$ lie close each to other. This may be in the
case of an accidental cancellation of the $Ry$ and $\alpha^2 Ry$
terms. An example of such a cancellation can be seen in the
spectrum of the Ag atom. Two excited multiplets, $4d^{10}5p$ and
$4d^95s^2$ (one of the $4d^95s^2$ lines is quite narrow $\sim 1$
Hz), are split slightly. The splitting comes from the gross
structure. However, it is comparable with the internal structure
of the multiplets, which is due to the relativistic corrections
(fine structure). For such a cancellation, a value of energy
splitting between levels from different multiplets is quite
sensitive to a variation of $\alpha$. Measuring the splitting with
relatively low accuracy, it is possible to reach a strong limit. A
similar idea for the $4f^{10}5d6s-4f^95d^26s$ lines in the $Dy$
atom was proposed in Ref. \cite{dzuba,dzuba99}. In some sense, the
search for the accidental degeneration is quite close to
approaches using nuclear data, in which the smallness of some
differences is  widely utilized (see e. g. Ref.
\cite{shlyakhter}).

Let us discuss conditions needed for success in such an
experiment. We consider the spectrum of the Ag atom  as an example
and some properties of low-lying levels in that atom are
presented in Fig. \ref{Fag} and Table \ref{Tag}.

\begin{figure}[h]
\epsfxsize=12cm
\centerline{\epsfbox{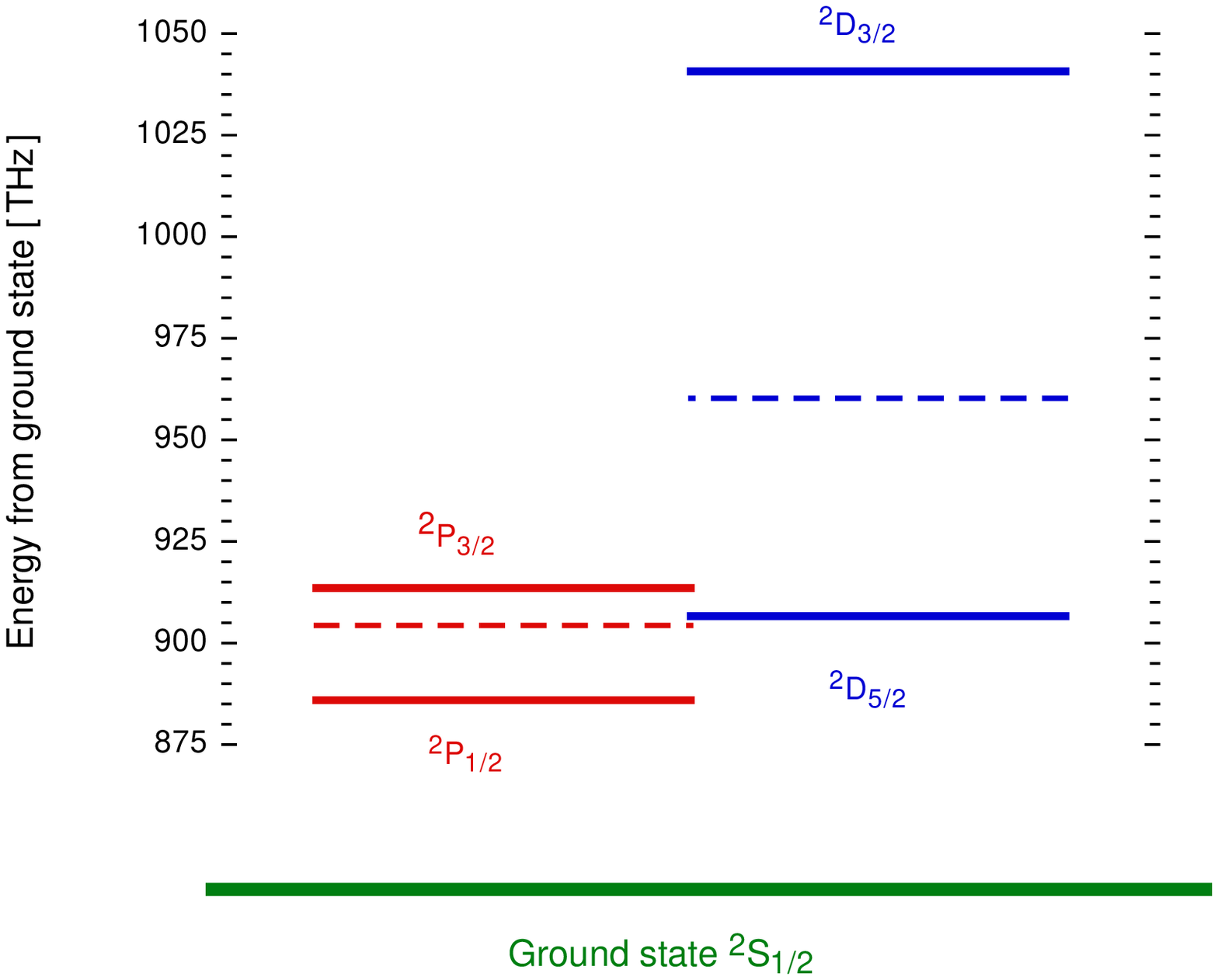}}
\caption{\label{Fag} \em Scheme of low-lying levels of the neutral Ag atom. The
hyperfine structure is not shown.}
\end{figure}

\begin{table}[th]
\begin{center}
\begin{tabular}{||c|c|c||}
\hline
\hline
 \multicolumn{3}{||c||}{} \\[-1ex]
 \multicolumn{3}{||c||}{Ground state: $4d^{10}5s\;{}^2S_{1/2}$} \\[1ex]
\hline
\hline
 \multicolumn{3}{||c||}{} \\[-1ex]
 \multicolumn{3}{||c||}{Excited states} \\[1ex]
\hline
&&          \\[-1ex]
Level                     &     Energy  & Lifetime   \\[1ex]
\hline
&&          \\[-1ex]
$4d^{10}5p\;{}^2P^0_{1/2}$  & ~885.94 THz & $8\cdot10^{-9}$ s  \\[1ex]
$4d^95s^2\;{}^2D_{5/2}$     & ~906.64 THz & 0.2 s             \\[1ex]
$4d^{10}5p\;{}^2P^0_{3/2}$  & ~913.55 THz & $7\cdot10^{-9}$ s  \\[1ex]
$4d^95s^2\;{}^2D_{3/2}$     & 1040.70 THz & $9\cdot10^{-5}$ s  \\[1ex]
\hline
 \multicolumn{3}{||c||}{} \\[-1ex]
 \multicolumn{3}{||c||}{Fine structure} \\[1ex]
\hline
&&          \\[-1ex]
Multiplet            &  The {\em fs}     & Center of      \\[1ex]
                     &  splitting        & gravity    \\[1ex]
\hline &&          \\[-1ex] $4d^{10}5p\;{}^2P^0_j$ & ~27.60 THz
&  904.35 THz     \\[1ex] $4d^95s^2\;{}^2D_j$    & 134.06 THz    &
960.24 THz     \\[1ex] \hline \hline
\end{tabular}
\caption{\label{Tag} \em Properties of low-lying excited states of the neutral
silver atom. The {\em hfs\/} is neglected.}
\end{center}
\end{table}

The conditions for a precise measurement of a transition sensitive
to possible variations of the constants are:
\begin{itemize}
\item The fine structure terms should be of the same order of
magnitude as the gross structure contributions. In the silver ion
this is true: the splitting of the two $5p{}^2P^0_J$ levels is
about the same value as the separation between one of them
($5p{}^2P^0_{3/2}$) and one of the $4d^95s^2{}^2D_J$ levels
($4d^95s^2{}^2D_{5/2}$). The other fine structure splitting
(between $D$ levels) is larger than the $P-D$ interval.
\item The non-relativistic ($Ry$) and relativistic terms
$\alpha^2 Ry$ have to have a different sign. This is very likely
because of the sandwich sequence of levels $P-D-P-D$ with the
center of gravity of the $D$ level lying above that for the
$P$ states. That means that the non-relativistic contribution for
the $D$ states is likely higher than for the $P$ states. As far as
we are interested in the lower $D$ level, we can expect that the
relativistic contribution to that is negative. Indeed, if silver
qualified for other conditions it would be studied theoretically
before performing the experiment. The relativistic effects shift
and split levels. We can estimate an enhancement of the
sensitivity assuming that there are no shifts, but only
splittings. We expect this should work for a preliminary
estimation. In such a case a non-relativistic correction can be
found from the separation of the centers of gravitation of the $P$
and $D$ lines:
\beq \Delta_{\rm NR} ({}^2P^0_{1/2}-{}^2D_{5/2}) = - 55.89~{\rm
THz} = A\,Ry\,,
\eeq
while the relativistic corrections are obtained from the shift of
the energy level from the position of the center of gravity
\beq
\Delta_{\rm Rel} ({}^2P^0_{3/2}-{}^2D_{5/2}) = 62.82~{\rm THz} = B\,\alpha^2\,Ry\,.
\eeq
The separation eventually is equal to
\beq
\Delta({}^2P^0_{3/2}-{}^2D_{5/2}) = 62.82~{\rm THz} - 55.89~{\rm THz} =
6.93~{\rm THz}\,.
\eeq
A variation of the Rydberg constant can be neglected and one can find
\beq
\frac{\partial \ln(\Delta/Ry)}{\partial \ln \alpha} \simeq
2 \,\frac{B\,\alpha^2 Ry}{\Delta}\approx 2\cdot 10\,.
\eeq
The factor 2 is the common factor because any relativistic correction depends on
$\alpha^2$ and 10 is an estimation of the
enhancement due to the accidental degenerations.
\item For accurate measurement it may be important to use laser
cooling. An important conditions for this is lack of the hyperfine
structure. Unfortunately both stable isotopes (${}^{107}$Ag and
${}^{109}$Ag) have some magnetic moment and the ground state of
any stable isotope is usually split into two states. The laser cooling
is possible but rather complicated.
\item It may also be important to apply two-photon Doppler-free
excitation to produce one of the two levels, splitting of which
contains a cancellation between relativistic and non-relativistic
term. This is necessary because it is somehow possible to cool the
ground state, but not excited states. We have to eliminate Doppler
effects due to excitation. The $D$ states in the silver atom can
be excited by means of the two-photon transitions.
\item Both levels have to be narrow. The $D$ level
($4d^95s^2{}^2D_{5/2}$) is very narrow with a width of about 1 Hz,
but both $P$ states are very broad.
\item A precise measurement of the small splitting due to the
cancellation must be possible. Generally this means that one must
be to induce a single-photon transition. The one-photon transition
between states $4d^95s^2{}^2D_{5/2}$ and $5p{}^2P^0_{3/2}$ lies at 7
THz. It can be induced but it is hard to measure such a transition
frequency precisely.
\end{itemize}
One can see that the conditions can be realized in some atomic
systems and it is necessary to search for them. One should note
that in the case of ions the condition for choice of nuclear spin
is different. It may be more important to have some states that
are insensitive to a magnetic field (the whole moment $F$ of a
system of electrons plus nucleus must be integer and states with
$m_F=0$ are alowed).

Our suggestion is to look for a relatively small enhancement but
with an appropriate possibility for precise measurements. The
proposal with $Dy$ in Refs. \cite{dzuba,dzuba99} is rather for a
great enhancement without any accurate spectroscopy.
\end{itemize}

\section{Nuclear magnetic moments and interpretation of the
frequency comparisons \label{inter}}

A frequency comparison involves atoms that are very differentin
nature and one must be prepared to interpret results. Most precise
results are for the {\em hfs\/} (see Table \ref{Trf}) and we
mainly discuss the {\em hfs\/} transitions. Since any absolute
measurements assume a comparison to the cesium hyperfine
separation this is also important for interpreting the absolute
optical measurements (see Table \ref{Topt}).

\subsection{Magnetic moments}

To the leading order the {\em hfs\/} interval can be presented in
the form of \eq{theohfs}. There are two different factors
important for the comparison: magnetic moments and the
relativistic corrections. The magnetic moments and some other
nuclear properties are collected in Table \ref{Tprop}. Some atoms,
hyperfine separation in which was measured accurately, but less
precisely than one part in $10^{11}$, are also included:
\begin{itemize}
\item $\nu_{\sc hfs}({}^{43}{\rm Ca}^+) = 15\,199\,862.858(2)$ kHz \cite{arbes};
\item $\nu_{\sc hfs}({}^{113}{\rm Cd}^+) = 3\,225\,608.286\,4(3)$ kHz \cite{tanaka};
\item $\nu_{\sc hfs}({}^{131}{\rm Ba}^+) = 9\,107\,913.699\,0(5)$ kHz \cite{knab};
\item $\nu_{\sc hfs}({}^{135}{\rm Ba}^+) = 7\,183\,340.234\,9(6)$ kHz \cite{becker83};
\item $\nu_{\sc hfs}({}^{137}{\rm Ba}^+) = 8\,037\,741.667\,7(4)$ kHz \cite{blatt82}.

\end{itemize}
Most of these are stable, expect cadmium ($T_{1/2}({}^{113}{\rm
Cd})=9\cdot 10^{15}$ yr) and the lightest barium
($T_{1/2}({}^{131}{\rm Ba})=12$ d). The sign of the magnetic
moment of $^{131}$Ba was presented in Ref. \cite{firestone} as
unknown and we follow Ref. \cite{knab}.

A discussion on the value of the nuclear magnetic moments is also
important because of our proposal to look for variations of small
moments.

{
%\small

\begin{table}[th]
\begin{center}
\begin{tabular}{||c|c|c|c||}
\hline\hline
&&&\\[-1ex]
    &           & Nuclear     & Magnetic        \\[1ex]
$Z$ &  Nucleus  & spin and    & moment          \\[1ex]
    &           & parity      & [$\mu_{\sc n}$] \\[1ex]
\hline
&&&\\[-1ex]
~  & H          & 1/2$^+$ & ~2.793 \\[1ex]
   & D          &   1$^+$ & ~0.857 \\[1ex]
   & T          & 1/2$^+$ & ~2.979 \\[1ex]
~4 & $^9$Be     & 3/2$^-$ & -1.178 \\[1ex]
20 & $^{43}$Ca  & 7/2$^-$ & -1.318 \\[1ex]
37 & $^{87}$Rb  & 3/2$^-$ & ~2.751 \\[1ex]
48 & $^{113}$Cd & 1/2$^+$ & -0.622 \\[1ex]
55 & $^{133}$Cs & 7/2$^+$ & ~2.582 \\[1ex]
56 & $^{131}$Ba & 1/2$^+$ & -0.708 \\[1ex]
   & $^{133}$Ba & 1/2$^+$ & -0.772 \\[1ex]
   & $^{135}$Ba & 3/2$^+$ & ~0.838 \\[1ex]
   & $^{137}$Ba & 3/2$^+$ & ~0.938 \\[1ex]
70 & $^{171}$Yb & 1/2$^-$ & ~0.494 \\[1ex]
   & $^{173}$Yb & 5/2$^-$ & -0.680 \\[1ex]
80 & $^{199}$Hg & 1/2$^-$ & ~0.506 \\[1ex]
\hline\hline
\end{tabular}
\vspace{5mm} \caption{\em Properties of some nuclei important for
precise microwave spectroscopy. \label{Tprop}}
\end{center}
\end{table}
}

All nuclei in Table \ref{Tprop} and \ref{Tmu} have an odd value of
$A$, while $Z$ is even for iron, gadolinium, osmium and tungsten
(Table \ref{Tmu}) and calcium, cadmium, barium, ytterbium and
mercury (Table  \ref{Tprop}) and odd for all others. An even value
of $Z$ indicates that the nuclear magnetic moment is associated
with the neutron magnetic moment and in the case of odd $Z$ the
moment is due to the proton one. Let us start with small moments.
Some of these can be understood using a simple model (the Schmidt
model), while assuming that the magnetic moment of the 
nucleus---like a moment of an electron in a hydrogen-like atom---includes
a spin part and an orbital part
\beq \label{schmidt}
\mbox{\boldmath{$\mu$}}_a(I) = \mu \,{\bf I}/{I}
= \mu_a^s {\bf S} + \mu_a^l {\bf L}\,,
\eeq
where ${\bf I}={\bf S} + {\bf L}$, $\mu_p^s=2\mu_p$,
$\mu_n^s=2\mu_n$, $\mu_p^l=\mu_{\sc n}$ and $\mu_n^l=0$. We should
remember that the values of the spin terms ($\mu_a^s$) originate
from dynamic effects of Quantum Chromodynamics (QCD) in the strong
coupling  regime and so they sensitive to a variation of the QCD
coupling constant ($\alpha_{\sc s}$). The Schmidt model leads to
some relatively small values in a few cases:
\begin{itemize}
\item for odd $Z$, $I=1/2$, $L=1$, in particular, N, Y, Rh and Ag
in Table \ref{Tmu} (a cancellation between spin and orbit
contributions)
\beq \label{p1/21}
\mu = \frac{4-g_p}{6}\,\mu_{\sc n} \sim -0.26 \,\mu_{\sc n}\,,
\eeq
where $g_p=2\cdot 2.793\dots$;
\item for odd $Z$, $I=3/2$, $L=2$, particularly, K, Ir and Au in
Table \ref{Tmu} (a cancellation between spin and orbit
contributions):
\beq  \label{p3/22}
\mu = \frac{3}{10}\,\Big(6-g_p\Big)\,\mu_{\sc n} \sim 0.12\, \mu_{\sc n}\,;
\eeq
\item for even $Z$, $I=1/2$, $L=1$, in particular, iron, osmium and
tungsten in Table \ref{Tmu}, ytterbium--171 and mercury in Table
\ref{Tprop} and $^{153}_{~62}$Sm, discussed in Sect. \ref{Shfs}
(the result is relatively small because of the coefficient 1/3):
\beq \label{HgYt}
\mu = -\frac{g_n}{6}\,\mu_{\sc n} \sim 0.64\, \mu_{\sc n}\,,
\eeq
where $g_n=-2\cdot 1.913\dots$.
\end{itemize}
In all other cases presented in Tables \ref{Tprop} and \ref{Tmu}
the value of the magnetic moment is not smaller than one in units
of the nuclear magneton $\mu_{\sc n}$. In some cases the agreement
between the Schmidt values and the actual ones is a 10\% level,
but in other cases the actual values are significantly smaller and
that is a result of the nuclear effects and, hence, small magnetic
moments are sensitive to these. Comparison of \eqs{p1/21},
\eqo{p3/22} and \eqo{HgYt} shows that the nuclei with odd $Z$ and
odd $A$ can be more interesting because the magnetic moment is
small partly due to cancellations between the spin and orbit
contributions, which are sensitive to variation of $g_p$.

We collect the Schmidt values for different nuclear spin
\beq \label{schmidt1}
\mu_{\sc s} (I=L\pm1/2) = I\, \left(g^l \pm \frac{g^s-g^l}{2l+1}\right)
\mu_{\sc n}\,,
\eeq
where $g^b=\mu^b /\mu_{\sc n}$ and $\mu^b$ for a proton and
neutron are defined above, in Table \ref{Tschmi}.

{
%\small

\begin{table}[th]
\begin{center}
\begin{tabular}{||c||c|c|c||c|c|c||}
\hline
\hline
& \multicolumn{6}{c||}{} \\[-1ex]
& \multicolumn{6}{c||}{Spin ($I$) and Schmidt value of the magnetic moment ($\mu_S$) 
of odd-$A$ nuclei} \\[1ex]
%& \multicolumn{6}{c||}{Magnetic moment ($\mu_S$) and spin ($I$) of odd-$A$ nuclei} \\[1ex]
\cline{2-7}
& \multicolumn{3}{c||}{} & \multicolumn{3}{c||}{} \\[-1ex]
$l$ & \multicolumn{3}{c||}{$I=l+1/2$} & \multicolumn{3}{c||}{$I=l-1/2$} \\[1ex]
\cline{2-7}
&  & \multicolumn{2}{c||}{} & & \multicolumn{2}{c||}{} \\[-1ex]
& $I$ & \multicolumn{2}{c||}{Magnetic moment [$\mu_{\sc n}$]} & $I$ & \multicolumn{2}{c||}
 {Magnetic moment [$\mu_{\sc n}$]} \\[1ex]
\cline{3-4}\cline{6-7}
& &  &  & & &  \\[-1ex]
& & Odd $Z$ & Even $Z$  & & Odd $Z$ & Even $Z$ \\[1ex]
\hline
& &  &  & & &  \\[-1ex]
0 & 1/2$^+$ & $g_p/2~~~~~\simeq 2.793$   & $g_n/2\simeq -1.913$      &
  -       & -  & -  \\[1.5ex]
1 & 3/2$^-$ & $g_p/2+1\simeq 3.793$ & $g_n/2\simeq -1.913$ & 1/2$^-$ &
  $1/3\,\big(2-g_p/2\big)\simeq - 0.264$ & $1/3\,\big(-g_n/2\big)\simeq 0.638$ \\[1.5ex]
2 & 5/2$^+$ & $g_p/2+2\simeq 4.793$ & $g_n/2\simeq -1.913$ & 3/2$^+$ &
  $3/5\,\big(3-g_p/2\big)\simeq ~0.124$   & $3/5\,\big(-g_n/2\big)\simeq 1.148$ \\[1.5ex]
3 & 7/2$^-$ & $g_p/2+3\simeq 5.793$ & $g_n/2\simeq -1.913$ & 5/2$^-$ &
  $5/7\,\big(4-g_p/2\big)\simeq ~0.862$   & $5/7\,\big(-g_n/2\big)\simeq 1.366$  \\[1.5ex]
4 & 9/2$^+$ & $g_p/2+4\simeq 6.793$ & $g_n/2\simeq -1.913$ & 7/2$^+$ &
  $7/9\,\big(5-g_p/2\big)\simeq ~1.717$   & $7/9\,\big(-g_n/2\big)\simeq 1.488$
\\[1.5ex]
\hline
\hline
\end{tabular}
\caption{\label{Tschmi} \em  Magnetic moment, spin and parity of nuclei in the Schmidt
model (\protect{\eqs{schmidt}} and \protect{\eqo{schmidt1}}).}
\end{center}
\end{table}
}

\begin{figure}[h]
\epsfxsize=16cm
\centerline{\epsfbox{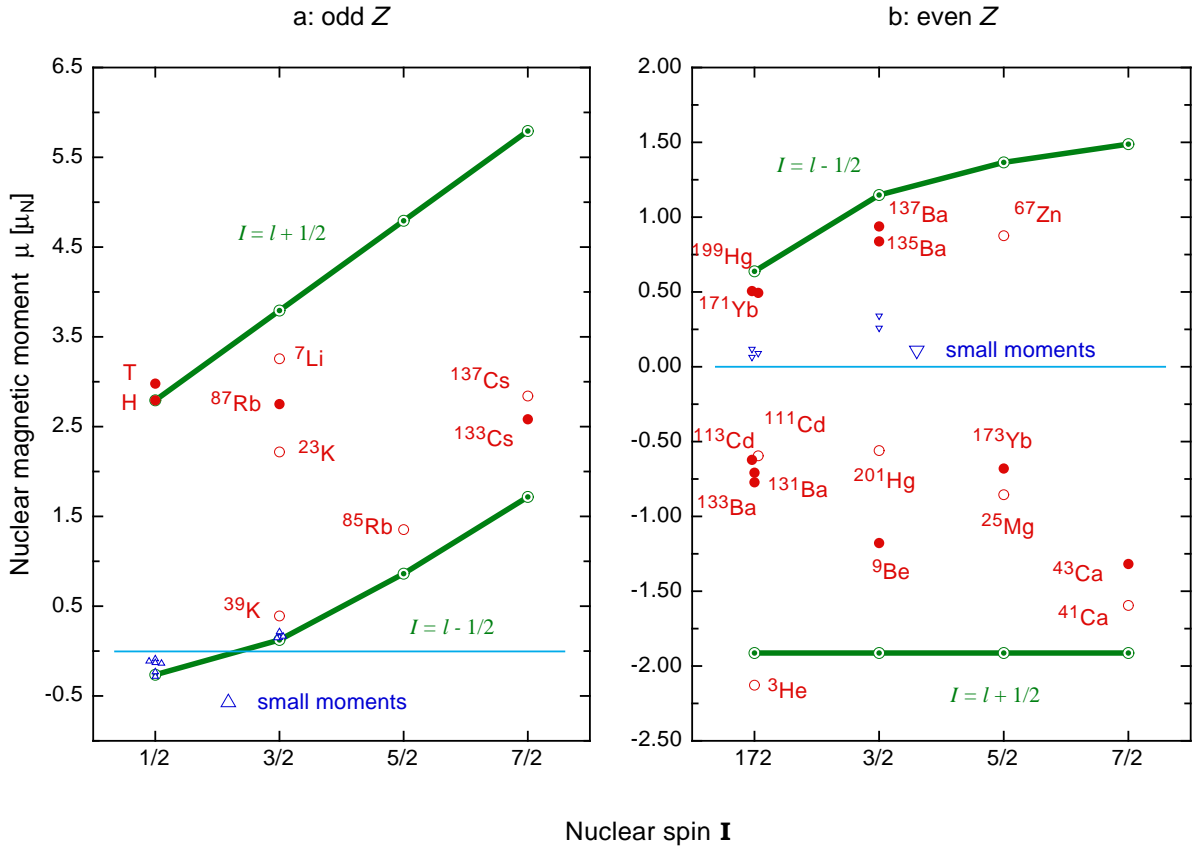}}
\caption{\label{Fsch} \em Magnetic moment of some isotopes: actual values and
the Schmidt model. Nuclei with small magnetic moments are listed in Table
\protect{\ref{Tmu}}.}
\end{figure}

Comparison of the Schmidt model with the most important isotopes
for precision measurements and variations of the constants is
presented in Fig. \ref{Fsch}. Here we collect the Schmidt values
(two lines) and actual values of the magnetic moments of the
isotopes from Table \ref{Tprop} (filled circles), other stable or
long-lived isotopes associated with simple atomic spectra of
neutral or single-charged ions (open circles), and the isotopes
from Table \ref{Tmu} with small magnetic moments (triangles). One
nucleus of this kind ($^{87}$Sr, 9/2$^+$, $\mu=-1.094\mu_{\sc n}$,
$\mu_{\sc s}(9/2^+)=-1.931\mu_{\sc n}$)) is not included in the
figure because of its large spin.

From the figure one sees that the agreement between real values
and the simple Schmidt model is not perfect. This means that
nature of nuclear spin are more complicated and the effects of
nuclear interaction are significant. However, the model contains
some important physics: e. g. the model predicts simple relations
between the nuclear parity and magnetic moment. Only one isotope
in Fig. \ref{Fsch} has inconsistent values of the parity and spin
($^{169}$Tm, 1/2$^+$, $\mu=-0.23\mu_{\sc n}$).

There is no simple model for even $A$, when there are two valence
particles---a proton and neutron---and a contribution of
orbital motion. For deuterium the proton and neutron contributions
have different signs \eqo{mud}. Two radionuclides discussed in
Sect. \ref{Shfs}, $^{198}_{~81}$Tl, $^{192}_{~79}$Au, have even
$A$.

\subsection{Relativistic corrections}

Now let us discuss the {\em hfs} intervals in Table \ref{Trf}.
First, we consider the relativistic corrections, which have been
in part discussed in Sect \ref{Clocom}. They can be calculated
within using many-body perturbation theory (see e. g. Ref.
\cite{dzuba99}). An approximation of \eqs{Fcasimir} and
\eqo{Lcasimir} is a good one for alkali atoms (all but ytterbium
and mercury in Table \ref{Trf}) under the conditions:
\begin{itemize}
\item High nuclear charge: $Z\gg 1$.
\item Atoms are neutral or slightly charged: $Z\gg z$, where $z$
is an effective charge of a compound nucleus for the valentce
electron, i. e. $z=1$ for neutral atoms and $z=2$ for
single-charged ones. Note that, \eq{Fcasimir} is incorrect for
hydrogen and its isotopes and for the Li-like ion of beryllium
presented in Table \ref{Trf}.
\item Accuracy is expected to be on a $5-15\%$ level and a
comparison of two atoms with a small charge difference ($Z_{1,2}
\gg |Z_1-Z_2|$) will not give accurate results.
\end{itemize}
As already mentioned, due to a discussion on the interpretation of
results in Ref. \cite{prestage}, any atom with a closed subshell
and one valence electron satisfies the above conditions. Alkali
atoms actually form only one example of such atomic systems.

Relativistic corrections are also important for the gross and fine
structure. They have a relative order of $(Z\alpha)^2$. In the
simplest case, namely for alkali atoms, they were discussed in
Ref. \cite{dzuba99}, where some results were obtained for Ca and
Sr$^+$ and in Ref. \cite{dzuba00} for In$^+$. It seems that
relativistic effects are less important for the optical
transitions than for the {\em hfs\/} because of relatively small
numerical coefficients and an extra $1/n^*$ factor.

In the hydrogen atom for both {\em hfs\/} and $1s-2s$ transitions
the corrections are negligible.

\subsection{Hyperfine structure}

The {\em hfs\/} of atoms with nuclei with even $Z$ and odd $A$ is
sensitive to the neutron magnetic moment. In particular the
Schmidt model predicts the magnetic moment of mercury and
ytterbium--171 (see \eq{HgYt}) within 15\% uncertainty. In
contrast the {\em hfs\/} of atoms with odd $Z$ and odd $A$ depends
on a value of the proton magnetic moment. An important point
however is the size of nuclear effects of the strong and
electromagnetic interactions. They can be estimated by a deviation
from the Schmidt value. In the case of Cs the nuclear effects
increase the value by a factor of about 1.5. This means that any
interpretation of comparisons with Cs cannot neglect the strong
interactions.

In the case of even-to-even or odd-to-odd comparisons, in
particular for H--Rb, and Be--Yb--Hg, we expect an information on
the variation of the constants due to the nuclear effects and the
relativistic corrections. Comparison of odd-to-even {\em hfs\/}
yields to variation of $g_n$ with respect to $g_p$. This is
important, because there is no reason to expect that a value of
$g_n/g_p$ is relatively stable, while the constants are varying.
Our point of view is different from that in Ref. \cite{prestage},
where the H--Hg$^+$ comparison was examined assuming that a
variation of the nuclear $g$-factors can be neglected.

We have not mentioned the Cs {\em hfs\/} because here the
interpretation is slightly different. Although ${}^{133}$Cs is an
isotope with odd $Z$, its comparison with H or Rb is sensitive to
a variation of $g_p$, because the $g_p$ contribution to the
nuclear $g$ factor has a negative sign in contrast to H and $Rb$.
For example, the sensitivity of the H-to-Cs comparison to the
$g_p$ variation is determined by the value of
\beq
\frac{\partial }{\partial \ln{g_p}}\,\ln\frac{\mu_{\sc s}({\rm Cs})}
{\mu_{\sc s}({\rm H})} = -\frac{10}{10-g_p} \simeq -2.3\,,
\eeq
and for the Rb-to-Cs comparison one can find
\beq \frac{\partial }{\partial \ln{g_p}}\,\ln\frac{\mu_{\sc
s}({\rm Cs})} {\mu_{\sc s}({\rm Rb})} = -\frac{12\,g_p}
{\big(2+g_p\big)\,\big(10-g_p\big)} \simeq -2.0\,.
\eeq
In contrast, the H-to-Rb comparison is actually insensitive to
any variations of the proton $g$-factor:
\beq
\frac{\partial }{\partial \ln{g_p}}\,\ln\frac{\mu_{\sc s}({\rm Rb})}
{\mu_{\sc s}({\rm H})} = -\frac{2}{2+g_p} \simeq -0.26\,.
\eeq

{
%\small

\begin{table}[th]
\begin{center}
\begin{tabular}{||c|c|c|c|c|c|c||}
\hline
\hline
& \multicolumn{4}{c|}{} &\multicolumn{2}{c||}{} \\[-1ex]
& \multicolumn{4}{c|}{Magnetic moment} &\multicolumn{2}{c||}{Relativistic} \\[1ex]
\cline{2-5}
&\multicolumn{2}{c|}{} &\multicolumn{2}{c|}{} &\multicolumn{2}{c||}{}\\[-1ex]
Atom  &\multicolumn{2}{c|}{Naive ($\mu_{\sc s}$)} &\multicolumn{2}{c|}{Actual ($\mu$)}
&\multicolumn{2}{c||}{correction}\\[1ex]
\cline{2-7}
 &  & & & & & \\[-1ex]
 & Eq. for $\mu_{\sc s}$ [$\mu_{\sc n}$] & [$\mu_{\sc n}$]
   & [$\mu_{\sc n}$] & $\mu/\mu_{\sc s}$ & $F(\alpha)$
&  $\partial \ln F/\partial \ln \alpha$         \\[1ex]
\hline
&&&&& &          \\[-1ex]
H              & $g_p/2$               & 2.79 & 2.79 & 1.00 & 1.00     & 0.00 \\[1ex]
D              & $(g_p-|g_n|)/2$       & 0.88 & 0.86 & 0.98 & 1.00     & 0.00 \\[1ex]
$^{9}$Be$^+$    & $g_n/2$             & -1.18 & -1.91 & 0.62 & 1.00 & 0.00 \\[1ex]
$^{87}$Rb      & $g_p/2+1$             & 3.73 & 2.75 & 0.74 & 1.15(10) & 0.30(6) \\[1ex]
$^{133}$Cs     & $\frac{7}{18}(10-g_p)$ & 1.72 & 2.58 & 1.50 & 1.39(7)
  & 0.83, \protect{\cite{dzuba99}} \\[1ex]
$^{171}$Yb$^+$   & $-g_n/6$              & 0.64 & 0.49 & 0.77 & 1.78 (9) & 1.42(15) \\[1ex]
$^{199}$Hg$^+$ & $-g_n/6$              & 0.64 & 0.51 & 0.80 & 2.26(12)
  & 2.30,  \protect{\cite{dzuba99}} \\[1ex]
\hline
\hline
\end{tabular}
\caption{\label{Tfin} \em Hyperfine structure properties.
Relativistic corrections, if not specified, are calculated from
\eqs{Fcasimir} and \eqo{Lcasimir}. We estimate the uncertainty for
$F$ as 5\% and for the derivative as 10\% by comparison with Ref.
\cite{dzuba99}. In the case of hydrogen and beryllium, the Casimir
correction is not appropriate, but \eqs{nr1}, \eqo{nr2} and
\eqo{nr3} lead to negligible shifts.}
\end{center}
\end{table}
}

We summarize some properties of {\em hfs\/} intervals of some
atoms in Table \ref{Tfin}. The naive value ($\mu_{\sc s}$) is the
Schmidt one ((see Table \ref{Tschmi}) for any atom, except
deuterium. The deuterium naive value is determined from \eq{mud}.
It is not quite clear if the Schmidt model works, but we expect it
to be appropriate for preliminary estimations and we consider
deviations from the naive value $\mu/\mu_{\sc s}-1$ as a
correction due to nuclear interactions. For Cs, the corrections
shift the value by 50\% and unsurprisingly that is the largest
contribution in Table \ref{Tfin}. Cesium has only one stable
isotope and this indicates that the nuclear configuration is not
strongly bounded and so different nuclear core polarization
effects or an admixture of excited states can be significant. We
have included the beryllium ion in the Table mainly due to the
measurement \cite{bollinger85} in the magnetic field. In the
leading non-relativistic approximation this value is proportional
to the {\em hfs} interval at zero field. The relativistic
corrections are different, but they are small enough.

Relativistic corrections are calculated using Casimir approach of
\eqs{Fcasimir} and \eqo{Lcasimir}, except for derivatives for Cs
and Hg, taken from Ref. \cite{dzuba99}. One reason, why
derivatives are relatively large is that the function $F$ is
always a function of $\alpha^2$. However, one can expect the same
for the strong interactions. One can see that the relativistic
corrections for alkali atoms are of the same sign and it must
be a partial cancellation for the comparison of two different {\em
hfs\/}. In contrast, the nuclear effects shift a value of
the nuclear magnetic moment in different directions for different
atoms and an enhancement (e. g. for the Rb--Cs {\em hfs\/}) is
possible. We can however hope that for a representative enough set
of atoms they can be considered rather as statistical errors.

Clearly if we would like to have a clear interpretation of some
frequency comparisons, we have to choose one of two options:
\begin{itemize}
\item We can measure only the gross and fine structure with
understandable relativistic effects and reach a limit for the
variation of $\alpha$.
\item We can involve the {\em hfs\/} in the comparison. In this
case we must be able to estimate nuclear effects and their
variations. The magnetic moment can be close to the Schmidt value
accidently due to some cancellation of contributions with
different nature and it can be quite sensitive to the variations
of the constants. So it is necessary to have small and
understandable corrections due to the nuclear effects. It is also
necessary to do a few comparisons in order to be able to transform
variations of the frequencies to variations of $\alpha$, $g_p$,
$g_n$ and $\mu_{\sc n}/\mu_{\sc b}$. One should emphasize a
significant difference between the study of atoms with odd $A$ and
either odd or even $Z$. The even $Z$ nuclear magnetic moments are
mainly proportional to the same value ($g_n$), which can be
corrected by nuclear effects. A comparison between two even $Z$
{\em hfs\/} is insensitive to any variations $g_n$. Conversely
investigation of only odd $Z$ atoms should provide enough
information on $g_p$ because they contain spin and orbit
contributions which diferently depend on nuclear quantum numbers
($L$, $S$ and $J$).
\end{itemize}

In principle, there is one more option search available. One can
look for two isotopes (with odd $Z$ and odd $A$, but different
nuclear spins or parity) of one element with magnetic moments
close to their Schmidt values. The ratio of the {\em hfs\/}
intervals should be free off any relativistic and many-body
atomic corrections and could be not too sensitive to nuclear
effects. In this case the ratio of the {\em hfs\/} should be close
to a ratio of their respective Schmidt values, i. e. a simple
function of $g_p$. Unfortunately we have been not able to find an
appropriate pair of the isotopes for this study.

\subsection{Time structure of the measurements}

We should point out a timing problem in the absolute frequency
measurements. Some results have been obtained via a direct
comparison to the primary cesium clock, i. e. directly compared to
the Cs {\em hfs\/}. Others have been compared indirectly and the
``time structure'' of such an experiment can be quite complicated
and includes some intermediate steps with comparisons between
different secondary standards. An example is a measurement of the
$2s-8s$ and $2s-8d$ transitions in hydrogen and deuterium atoms by
the Paris group. The measurements \cite{beauvoir} involved a
comparison in 1997 of the hydrogen and deuterium lines with some
secondary standard, gauged in 1985 \cite{clairon}. The standard
was recalibrated in 1999 \cite{rovera}. Such a time structure is
reasonable for a determination of the Rydberg constant and the
hydrogen Lamb shift, but it is not appropriate to search for a
variation of the constants. In Tables \ref{Trf} and \ref{Topt} we
assume that publication time is the time for a direct comparison
with a primary Cs standard. For the hydrogen and deuterium
hyperfine structure that is approximately correct. In some cases
that is not so.

The other time structure problem is in a comparison with secondary
standards, like a hydrogen maser. Some preliminary studies
estimated possible deviations of its frequency with respect to
some known standards or with respect to some average value of an
ensemble of masers. In both case it is not clear how to interpret
any comparison with a specific maser at some particular time.

\section{The hyperfine separation in the hydrogen atom}

\subsection{Historical remarks \label{Histo}}

The hydrogen {\em hfs\/} was studied for a few generations of
experiments. The first results with an accuracy of about
$10^{12}$ were reached about 35 years ago and some of them are
presented in Table \ref{Thist}. Most experiments
\cite{vessot,becker,mungall,chi} were devoted to a measurement of
the maser frequency, while previously measured value for the
wall-shift using another maser in Ref. \cite{vanier} was
accepted. In Ref. \cite{mungall} no explicit value of uncertainty
was claimed. From discussions in that paper we estimate it as
$3\cdot10^{-12}$ for the NRC cesium standard and
$1.7\cdot10^{-12}$ for the maser.

{
%\small

\begin{table}[th]
\begin{center}
\begin{tabular}{||c|c|c|c|c||}
\hline\hline
&&&&\\[-1ex]
 Frequency ($\nu$) & Ref. to   & Year & Relative    & Ref. to \\[1ex]
           [kHz]   & frequency &      & uncertainty & wall-shift   \\[1ex]
\hline
&&&&\\[-1ex]
1 420 405.751 786(2) &   \protect{\cite{vessot}}, 1966 & 1965
   & $1.4 \cdot 10^{-12}$  & \protect{\cite{vanier70}} \\[1ex]
1 420 405.751 756(3) &   \protect{\cite{becker}}, 1968 & 1966
   & $2.1 \cdot 10^{-12}$  & \protect{\cite{vanier70}} \\[1ex]
1 420 405.751 758(2) &   \protect{\cite{becker}}, 1969 & 1967
   & $1.4 \cdot 10^{-12}$  & \protect{\cite{vanier70}} \\[1ex]
1 420 405.751 776(5) &   \protect{\cite{mungall}}, 1968 & 1968
   & $3.5 \cdot 10^{-12}$  & \protect{\cite{vanier70}} \\[1ex]
1 420 405.751 777(3) &   \protect{\cite{chi}}, 1970  & 1968
   & $2.1 \cdot 10^{-12}$  & \protect{\cite{vanier70}}\\[1ex]
\hline\hline
\end{tabular}
\vspace{5mm} \caption{\em Some early precise measurements of the
hydrogen hyperfine separation. The references are given with the
year of publication, while the year of the experiment is given
additionally. \label{Thist}}
\end{center}
\end{table}
}

\begin{figure}[h]
\epsfxsize=17cm \centerline{\epsfbox{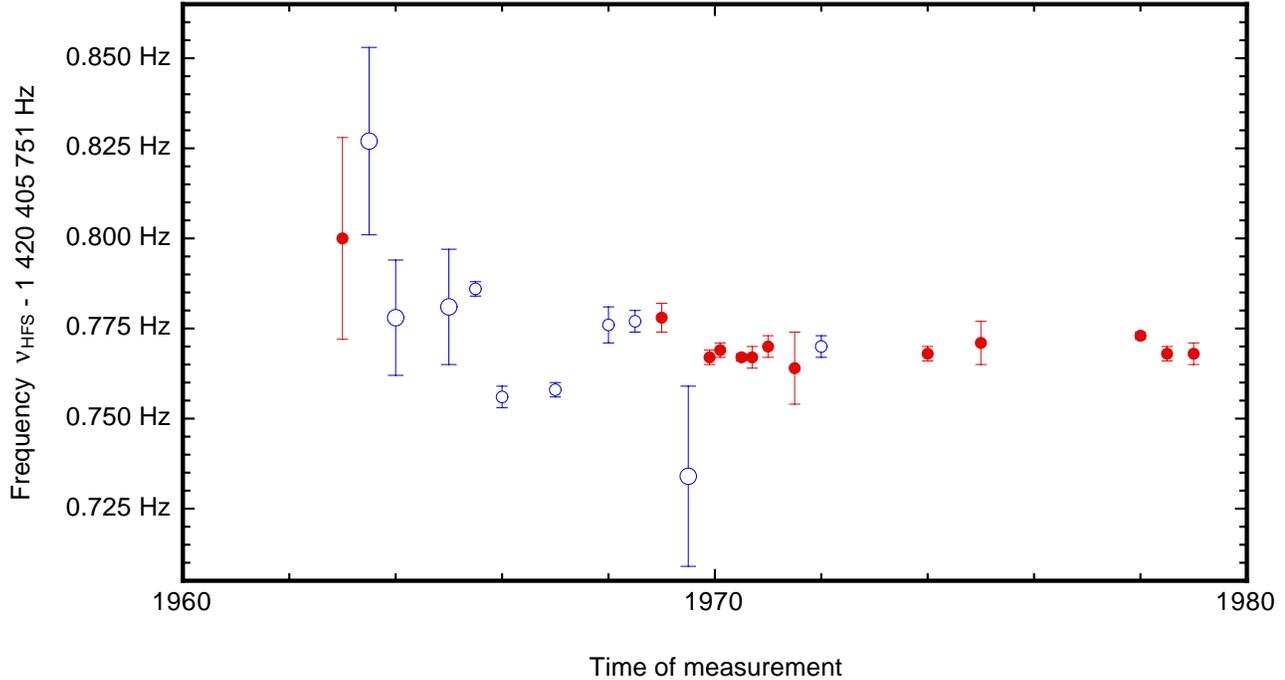}}
\caption{\label{Fall} \em Precise experimental results on the
hydrogen hyperfine separation. Filled circles are for those when the
wall-shift was measured together with the frequency, while the
open ones are for frequency measurements using a value of the
wall-shift found separately for another maser.}
\end{figure}

One of the first really accurate experiments was performed by NBS
\cite{hellwig} and in part by Harvard University team
\cite{zitzewitz}. It was pointed out \cite{hellwig} that the
wall-shift and the frequency have to be determined in the
experiments for the same masers. This generation of experiments
\cite{menoud,hellwig,zitzewitz,essen,morris,essen73,reinhard,petit,vanier}
is discussed in the next section. When we speak about two
generations we refer to an ideology, rather than to a time-frame.
Some other experiments
\cite{crampton,vanier64,peters1,peters2,johnson,bangham} performed
in that time or slightly earlier were not so precise (see Fig.
\ref{Fall}). Let us mention Ref. \cite{crampton} wherein an
experiment that measured both the frequency and the wall-shift
was described. However, only two bulbs were used. As was noted in
Ref. \cite{hellwig} another important condition for appropriate
results is the use of a large number of bulbs of different size
(e. g. in Ref. \cite{hellwig} 11 bulbs were and in Ref.
\cite{zitzewitz} the number was 18).

\subsection{Thirty years ago \label{Hhfs}}

About thirty years ago a number of precise results for the {\em
hfs} interval in the ground state of the hydrogen atom were
published
\cite{menoud,hellwig,essen,morris,essen73,reinhard,petit,vanier,cheng,petit80}.
That was due to a trial to use the hydrogen maser as a primary
frequency standard. The results are collected in Table
\ref{Thhfs}. Until the publication of results \eq{beplus} of
experiment on the {\em hfs\/} of the beryllium ion fifteen years
ago \cite{bollinger85}, the value of the hyperfine separation in
the hydrogen atom had been the most precisely measured physical
quantity.

{
%\small

\begin{table}[th]
\begin{center}
\begin{tabular}{||c|c|c|c|c|c|c||}
\hline\hline
&&&&&&\\[-1ex]
\# & Frequency ($\nu$) & Ref.  & Year & Relative & Cs standard & Comment \\[1ex]
      &      [kHz]& & & uncertainty &   &\\[1ex]
\hline
&&&&&&\\[-1ex]
~1 & 1 420 405.751 778(4) &   \protect{\cite{menoud}}, 1969 &  1969 &
     $28 \cdot 10^{-13}$ & LSRH commerc.  & 2 bulbs \\[1ex]
~2 & 1 420 405.751 769(2) &   \protect{\cite{hellwig}}, 1970 & 1969--1970 &
     $14 \cdot 10^{-13}$ & NBS primary & Exp. 1, 12 bulbs \protect{\cite{zitzewitz}} \\[1ex]
~3 & 1 420 405.751 767(2) &   \protect{\cite{hellwig}}, 1970 & 1969--1970 &
     $14 \cdot 10^{-13}$ & NBS primary & Exp. 2, 18 bulbs \\[1ex]
~4 & 1 420 405.751 768(2) &   \protect{\cite{hellwig}}, 1970 & 1969--1970 &
     $14 \cdot 10^{-13}$ & NBS primary & Exp. 1 \& 2\\[1ex]
~5 & 1 420 405.751 767(1) & \protect{\cite{essen}}, 1971 &  1970  &
     $7.0 \cdot 10^{-13}$ & NPL primary & 6 bulbs \\[1ex]
~6 & 1 420 405.751 770(3) & \protect{\cite{morris}}, 1971 & 1970--1971 &
     $21 \cdot 10^{-13}$ & NRC primary & 5 bulbs \\[1ex]
~7 & 1 420 405.751 767(3) & \protect{\cite{essen73}}, 1973 &  1970 &
     $21 \cdot 10^{-13}$ & NPL primary &  6 bulbs\\[1ex]
~8 & 1 420 405.751 768(2) &  \protect{\cite{reinhard}}, 1974 & 1974 &
     $14 \cdot 10^{-13}$ & LORAN C, USNO  & Flexible bulb \\[1ex]
~9 & 1 420 405.751 770(3) & \protect{\cite{petit}}, 1974 & 1972  &
     $21 \cdot 10^{-13}$ & TOP, TAF & Wall-shift \protect{\cite{zitzewitz}} \\[1ex]
10 & 1 420 405.751 771(6) &   \protect{\cite{vanier}}, 1978 & 1975--1976  &
     $41 \cdot 10^{-13}$ & LORAN C & 6 bulbs \\[1ex]
11 & 1 420 405.751 768(2) &   \protect{\cite{cheng}}, 1980 & 1979  &
     $14 \cdot 10^{-13}$ & LORAN C & 5 bulbs \\[1ex]
12 & 1 420 405.751 768(3) &   \protect{\cite{cheng}}, 1980 & 1979  &
     $21 \cdot 10^{-13}$ & LORAN C & 5 bulbs \\[1ex]
13 & 1 420 405.751 773(1) &   \protect{\cite{petit80}}, 1980 & 1978  &
     $~7 \cdot 10^{-13}$ & TAF &  Flexible bulb, \\[1ex]
   &                     &                                   &       &
                         &     &  zero wall-shift \\[1ex]
\hline\hline
\end{tabular}
\vspace{5mm} \caption{\em The most precise measurements of the
hydrogen hyperfine structure interval. The references are given
with the year of publication, while the year of the experiment is
given additionally. The abbreviations are: {\em \/} 
{\em NBS\/}---former National 
Bureau of Standards (USA), {\em NPL\/}---National
Physical Laboratory (UK), {\em NRC\/}---National Research Council
(Canada), {\em LORAN C\/}---a navigation system signal controlled
by a cesium standard, {\em USNO\/}---U. S. Naval Observatory,
{\em TOP\/}---cesium clock from the Paris Observatory, {\em
TAF\/}---the French atomic time.
%{\em TAI\/}---international atomic time.
\label{Thhfs}}
\end{center}
\end{table}
}

The experiments were performed using hydrogen masers and two key
values were measured simultaneously: the frequency of the maser
and the wall-shift. Result \#9 is an exception: the authors
utilized a value of the wall-shift from Ref. \cite{zitzewitz}.
Result \#5 of Ref. \cite{essen} with the smallest uncertainty was
actually a preliminary presentation. The final value of the NPL
experiment (result \#5) had an accuracy \cite{essen73} three
times lower. One of the papers contains a report on two
independent measurements \cite{hellwig}, quoted as experiment 1
and experiment 2, and we give both results in the table, as well
as an average value presented by the authors of Ref.
\cite{hellwig}. Experiment 2 is a pure NBS experiment. In
experiment 1 the maser frequency of a maser from Harvard
university was measured with respect to the NBS primary cesium
standard \cite{hellwig}, while the wall-shift of the same maser
was measured by the Harvard people \cite{zitzewitz}. The same
result for the wall-shift \cite{zitzewitz} was accepted by the
authors of Ref. \cite{petit} for another maser, the frequency of
which they measured. Values \#11 and \#12 were presented in Ref.
\cite{cheng}. That paper is devoted to a measurement of the
wall-shift and the unperturbed maser frequency (result \#11) at
the Shanghai Bureau of Metrology. It also contains a reference to
an unpublished result (\#12) for the unperturbed hydrogen
frequency of two hydrogen maser by the Shaanxi Asrtonomical
Observatory. In this two bulbs from the SBM were used. The last
result in Table (\# 13) is based on a different idea. Authors
studied the temperature dependence of the wall-shift and they found
the the shift vanished at some temperature $T_0$. The {\em hfs} interval
was afterwards determined from a maser frequency at $T_0$.

The value in Table \ref{Trf} (\eq{average}) calculated by Ramsey
\cite{ramsey} is an average of results \#4 and \#5, and so cannot
be actually accepted. However, the results  in Table \ref{Thhfs}
are consistent (apart from the earliest one) and one can calculate
some average values. For instance the average values can be
calculated over results \#\#1--3, 6--13 (a {\em wide\/} set) or
over \#\#4, 6, 7, 10 and 11 (a {\em conservative\/} set). The wide
set includes all original results, while the conservative one
contains only values published in refereed journals. We consider
results \#1 and \#7 as preliminary and also exclude result \#8
because the wall-shift and the frequency werenot  measured during
the same experiment.

\begin{figure}[h]
\epsfxsize=17cm \centerline{\epsfbox{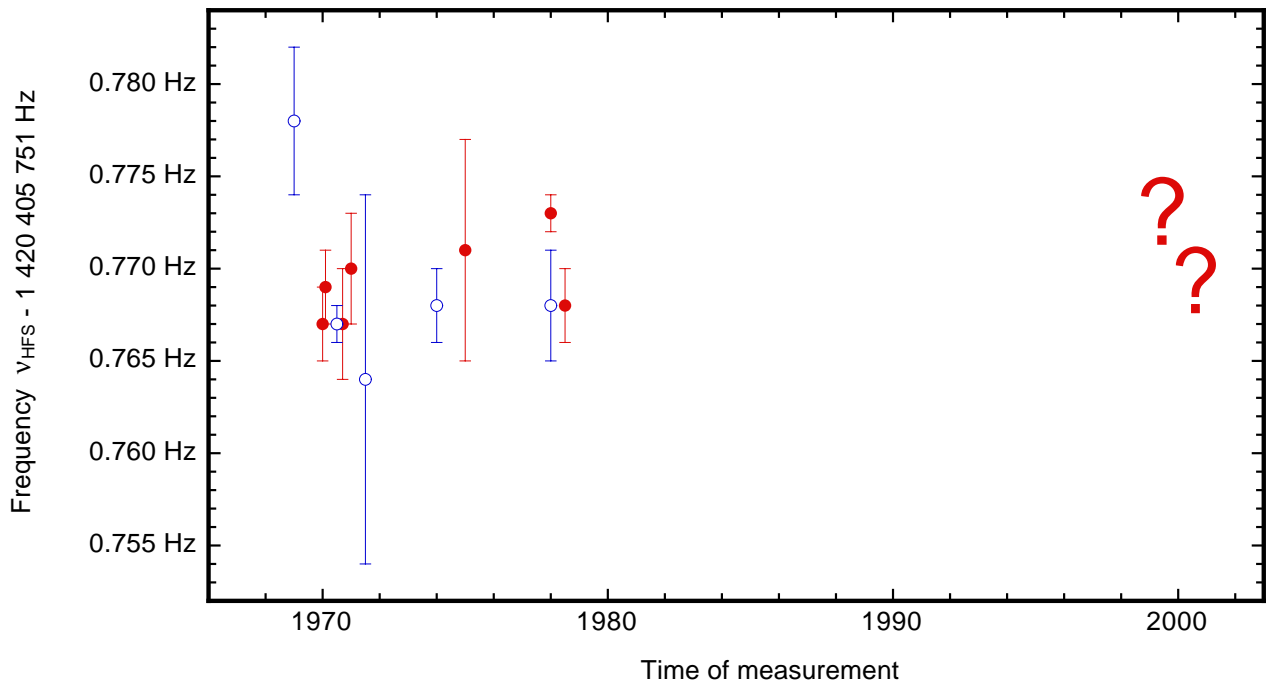}}
\caption{\label{Fbest} \em Hydrogen hyperfine structure as
measured thirty years ago. Result with measurement of the
wall-shift and the frequency for the same masers. Open circles
are for preliminary results (conference proceedings, private
communications) and results, corrected after publications. Full
circles are for the final results. The question marks are for the
possible 2000 measurements.}
\end{figure}

As a preliminary estimation, one can assume that the uncertainties
are independent and one finds:
\beq \label{wide}
\nu_{\rm wide} =1\,420\,405.751\,770\,4(6)\;{\rm
kHz}
\eeq
and
\beq \label{cons}
\nu_{\rm cons} = 1\,420\,405.751\,768\,3(12)\;{\rm kHz}\,.
\eeq
The fractional uncertainty of the average value is (4--9)$\cdot
10^{-13}$ and that is approximately as in Table \ref{Trf}. In fact
there has to be some correlation between the systematic errors.
However since most of the results are consistent, any averaging
cannot yield a value that is less precise than one of the results,
namely
\begin{equation} \label{nbs}
\nu(\#4)=1420\, 405.751\, 768(2)\; {\rm kHz}\,.
\end{equation}
We think this result should be used for comparisons because it is
the most accurate and reliable result in Table \ref{Thhfs}. Result
\#4 was published in a refereed journal \cite{hellwig} (in
contrast to result \#8 \cite{reinhard}) and it was based on two
independent measurements of the unperturbated hydrogen {\em hfs\/}
frequency with respect to the NBS primary Cs standard. The largest
number of bulbs was used to determine the wall-shift using two
independent experiments \cite{zitzewitz,hellwig}. The result is in
fair agreement with other results in Table \ref{Thhfs}, apart from
results \#1 and \# 13. The discrepancy with the former is not
important, because result \#1 does not agree with others on a
one-sigma level, but rather within three sigma. We also should
mention that Ref. \cite{morris} refers to a private communication
by one of the authors of Ref. \cite{menoud} (Menoud, 1971) wherein
a new LSRH value of hydrogen {\em hfs\/}
\beq
\nu({\em LSRH})=1\,420\,405.751\,764(10)~{\rm kHz}\,,
\eeq
is quoted. This is less precise than result \#1 but agrees with
other results. Two results in Table 15 are more precise than the
value in \eq{nbs}. One of them (\# 5) \cite{essen} was later
corrected by the authors \cite{essen73}. The other (result \# 13)
was obtained using very different methods. We do not include it in
the conservative set since it is inconsistent with most other
results of this set. To conclude, we present in Fig. \ref{Fbest}
an overview of experiments done about 30 years ago. Most
experimental results measured the wall-shift together with the
frequency. Some details on the crucial results can be found in the
Appendix.

One should note that 30 years ago the accuracy of the primary
cesium standards was not so high as now and this was an important
source of uncertainty. The results in Table \ref{Thhfs} are
presented together with the name of the primary cesium standard
used. We expect that it is possible to study variations of those
standards which were frequently discussed due to international
comparisons. We also expect that the reference set for comparison
with experiments possible in the year 2000 will to be between the
{\em conservative\/} and {\em wide\/} ones. Particularly, the main
sources of uncertainty in experiments 1 and 2 in Ref.
\cite{hellwig} are, in part, independent.

\subsection{Why it is important to do experiment with hydrogen now}

The hydrogen atom can now be used in the search for a limit for
the variations of fundamental constants on a level that is
comparable with most other possible projects.  If the Paris group
have really reached a level of accuracy ($\delta\nu/\nu\simeq
2.4\cdot 10^{-15}$) reported \cite{laurent} as a preliminary
result
\beq \nu_{\sc hfs}({}^{87}{\rm Rb}) =
6\,834\,682.610\,904\,333(17)\;{\rm kHz}\,,
\eeq
then only the work on the Rb clock and the Rb {\em hfs\/} can
provide a significantly more precise test for the variations.
Usually, it takes some time to really remove all sources of
systematic errors after the first announcement of a performed
measurement with a significantly better accuracy. The ratio of Rb
and Cs hfs may also be less sensitive to the variations, and
therefore it is not enough to compare two {\em hfs\/} transitions
only. It is necessary to have several different data sets.

In any case, the study of the hydrogen {\em hfs\/} has a number of
advantages, as are listed below:
\begin{itemize}
\item This is the only accurate comparison over a long interval
(30 yr). Only the $^{171}$Yb$^+$ {\em hfs\/} and the $^9$Be$^+$ in
the magnetic field (with a shorter time separation) is comparable
to this. The other results obtained five or more years ago cannot
provide a search for a variation of the constants on a level below
$10^{-13}\,{\rm yr}^{-1}$.
\item The result was obtained 30 years ago  independently in a
few different laboratories (see Table \ref{Thhfs}) after long
studies over a number of years. Only ytterbium--171 was measured
in two laboratories, whereas other transitions from Tables
\ref{Trf} and \ref{Topt} have been studied in only one place for
each atom.
\item Most of the measurements of the hydrogen {\em hfs\/} were
performed by direct comparisons with primary Cs clocks and thus
they have simple time structures.
\item Nowadays, this measurement can be also done in a number of
laboratories.
\item There is no direct problem due to the accuracy of
any primary frequency standard because of the relatively low
accuracy needed for the measurement.
\item The measurement should be as precise as possible. If a
level better than 10$^{-14}$ is achieved, a repetition in a few
years can yield another strict limit for the variation of the
constant. Thus the measurement in the year 2000 can be the
``second'' measurement of a 30-years test and the ``first''
measurement of a few-years test of the variation.
\item Combining the variation of the H--Cs {\em hfs\/} with
a variation of $1s-2s$ in the hydrogen atom with respect to the Cs
{\em hfs\/}, one can achieve an estimation for a variation of the
hydrogen {\em hfs} with respect to the Rydberg constant. The
former limit is going to be about $(2-5)\cdot 10^{-14}$ yr$^{-1}$.
The accuracy of a determination of the $1s-2s$ frequency in the
hydrogen atom with respect to the Cs {\em hfs\/} has been improved
and the uncertainty in 1999 \cite{1999} is about $6\cdot 10^{-14}$
yr$^{-1}$. This indirect comparison between the {\em hfs\/} and
the $1s-2s$ frequency in hydrogen can be interpreted without any
difficulties arising from the relativistic corrections or unclear
nuclear effects. Only properties of fundamental particles are to
be compared: $\alpha$ and $\mu_p/\mu_{\sc b}$.
\end{itemize}

The hydrogen experiment must have systematic errors, which are
quite different from the short-term comparisons. The value of
$\mu_p$ is on for the simplest for interpreting. The result can be
reproduced in different laboratories. The expected limit for the
variation of the hydrogen {\em hfs\/} lies between
$2\cdot10^{-14}$ yr$^{-1}$ and $5\cdot10^{-14}$ yr$^{-1}$. This
depends on the possibility of averaging the values from Table
\ref{Thhfs}. Some additional analysis of the old data
\cite{menoud,hellwig,morris,essen73,reinhard,petit,zitzewitz,vanier}
is needed. This would include the more recent data of
international comparisons on the national primary standards
utilized in the hydrogen experiments
\cite{menoud,hellwig,morris,essen73,reinhard,petit,vanier}. In
2000 only the Rb {\em hfs\/} can provide a significantly better
result. From published values, we estimate the limits from
possible experiments to approximately $1.3\cdot10^{-14}$ yr$^{-1}$
(for Rb) and $1.1\cdot10^{-14}$ yr$^{-1}$ (for Hg$^+$). The latter
is different from that presented in Table \ref{Trf} because we
expect significant correlations between the uncertainties of the
original measurement and the repetition.

\section{Conclusions}

The strongest published limits (better than $10^{-16}$ yr$^{-1}$)
on the variation of fundamental constants arise from geophysical
data. However, any direct use of such estimations is not possible
because of a incorrect interpretation of the data. The best
astrophysical data gives limits between $10^{-14}$ and $10^{-15}$
yr$^{-1}$ and they are only slightly better than laboratory
limits. Cosmological methods (such as investigation of
nucleosynthesis and microwave background radiation) are far less
precise, but they may be important if the variations at the
beginning of the evolution of the universe were faster. The clock
comparisons lead to limit on a level between $10^{-13}$ and
$10^{-14}$ yr$^{-1}$, although, their interpretation is not quite
clear. We summarize the different methods used to search for the
variations in Table \ref{Tmeth}. The data contain the amplitude
and velocity of the variations and the time and space separations
where appropriate. The scale of the space variations due to
geochemical studies is estimated from the absolute motion of the
Earth (i. e. a motion with respect to the frame of the microwave
background radiation frame).

\begin{table}[th]
\begin{center}
\begin{tabular}{||c|c|c|c|c||}
\hline\hline
&&&&\\[-1ex]
Method & $\Delta \ln\alpha/\Delta t$ & $\Delta\alpha/\alpha$
  & $\Delta t$ & $\Delta l/c$ \\[1ex]
\hline &&&&\\[-1ex]
Geochemical (Oklo)  & $10^{-17}~ {\rm
yr^{-1}}$ & $10^{-8}$
  & $2\cdot 10^{9}~ {\rm yr}$ & $10^6$~ yr \\[1ex]
Astrophysical & $10^{-15}~ {\rm yr^{-1}}$ & $10^{-5}$
  & $ 10^{9}-10^{10}~ {\rm yr}$ &  $ 10^{9}-10^{10}~ {\rm yr}$ \\[1ex]
Cosmological  & $10^{-13} {\rm~ yr^{-1}}$ & $10^{-3}$
  & $10^{10}$ yr & $10^{10}$~ yr \\[1ex]
Laboratory    & $10^{-14} {\rm~ yr^{-1}}$ & $10^{-14}$
  & $1$~ yr &  \\[1ex]
H/Cs {\em hfs\/}   & $(2-5)\cdot10^{-14} {\rm~ yr^{-1}}$ & $(6-15)\cdot10^{-13}$
  & $30$~ yr &  \\[1ex]
\hline
\hline
\end{tabular}
%\vspace{5mm}
\caption{\label{Tmeth} \em Comparison of different search for the variations.}
\end{center}
\end{table}

We believe that the most reliable limits can be reached under
laboratory conditions by comparing two results for the same
transition, each result obtained in different time. In 2000, there
are a number of possibilities to reach results a limit of a few
units of $10^{-14}$ yr$^{-1}$. The most secure one will be that
involving the hydrogen hyperfine splitting. The limit for a
possible variation of the ratio of $\mu_{\rm Cs}/\mu_p$ is
expected to be about $(1-2)\cdot10^{-14}$ yr$^{-1}$. The
$\alpha$-variation due to the relativistic corrections is to be
limited by $(2-5)\cdot10^{-14}$ yr$^{-1}$. The ytterbium limit for
the variation of $g_n/g_p$ is expected to be $6.5\cdot10^{-14}$
yr$^{-1}$. We need to mention, that the preliminary results on the
second ytterbium experiment were in fact published in 1995, but
these were later corrected \cite{fisk}. A significant part of the
measurements was performed by the end of 1995, while the remainder
was done in mid 1996. If we accept that the experiment was done in
1995, the potential limit is $4\cdot10^{-14}$ yr$^{-1}$. For the
deuterium {\em hfs}, if we assume a variation of $g_p$ only, the
limit is about $5.5\cdot10^{-14}$ yr$^{-1}$, whereas if we only
consider $g_n$-variations it is $8.1\cdot10^{-14}$ yr$^{-1}$. The
motivation to study the deuterium {\em hfs\/} is that it is the
only value is sensitive to $g_p-|g_n|$, amongst those known for a
while in Table \ref{Trf}.

In Table \ref{Tmeth} one can note that the astrophysical,
cosmological and geochemical data are not quite sensitive to any
fluctuations of the constants in $T\sim 10^8$ yr and/or $L\sim
c\cdot 10^8$ yr. Astrophysical data yield good limits for the
value of $\Delta \ln{\alpha}/ \Delta z$, but any interpretation of
the redshift $z$ in terms of the time separation such as e.~g.
\beq
\Delta t=t_0 \left(1-\frac{1}{(1+z)^{3/2}}\right)
\,,
\eeq
where $t_0=1.5\cdot 10^{10}$ yr, is actually only an estimation.
It is known that the Hubble velocity-distance law is not a strong
one and this means that for any particular case, there is no
well-established connection between the redshift (associated with
the velocity) and the time $t(z)$ (associated with the distance).
This also means that would even be hard to detect a fluctuation
with $T\leq 10^9$ yr. Since the interpretation is actually based
on a lack of such fluctuations the astrophysical and geochemical
limits are actually weaker than presented in the table. Another
problem is the correlation between space and time variations. If
we suppose e. g. that the constants increase with time and
distance, we have to expect a significant cancellation between the
time and space variations (actually we only study some distant
objects from the past at a distance $\Delta l=c\,\Delta t$).
Thence the astrophysical data can be insensitive to some of thse
correlations. It must also be mentioned that some particular
astrophysical data rather indicate in existence of some variations
of the constants. E. g. a few points in Fig. 1 of Ref. \cite{webb}
confirm a variation in $\alpha$. However, there are a lot of data
points plotted and, statistically, perhaps those few are not
important. nevertheless, the fit assumes that there is only a slow
drift in time and that there are neither space variations and nor
fluctuations on a scale shorter than $10^{10}$ yr. We think these
points need to be re-examined in order to understand if the
effect is purely statistical.

In contrast to examinations of the astrophysical and geochemical
data, a laboratory experiment can determine the derivatives of the
constants and is also sensitive to the fluctuations. There should
also be no cancellation between space and time deviations: no
space variation is involved because of the small absolute velocity
of Earth.

There are four basic dimensionless constants which determine any
atomic spectra (in unit of $Ry$): $\alpha$, $m_e/m_p$, $g_p$ and
$g_n$.  When the frequency is measured in units of the Rydberg
constant, the nuclear magnetic moment comes in units of the Bohr
magneton $\mu_{\sc b}$. Some of these constants enter into the
equations for the energy levels with nuclear magnetic moments
constructed of the proton spin contribution
$\mu_p=g_p\,m_e/m_p\,\mu_{\sc b}$, the neutron (spin) contribution
$\mu_n=g_n\,m_e/m_p\,\mu_{\sc b}$ and the proton orbital
contribution $m_e/m_p\,\mu_{\sc b}$. The last combination can also
appear as a Dirac part of the proton spin contribution for
relativistic corrections. These are known to split the Dirac part
and the anomalous part of the magnetic moment. This simple,
four-constant description is perturbed by nuclear effects.

We would also like to underline that there are two kinds of
searches for the variations. One is for clear interpretation and
it involves transitions without the influence of nuclear effects
or with only a small influence. This is valid for secure limits on
the possible variations of the constants. The other kind involves
studying transitions which may be expected to be extremely
sensitive to some variations, but for which there in no clear
interpretation. Particularly, the search for sensitive values
includes some geochemical study (positions of low-lying
resonances) and some laboratory investigations.

Astrophysical studies are appropriate in the search for variations
of $\alpha$, $m_e/m_p$ and $g_p$. The limits for $g_n$ cannot be
obtained from astrophysics. If nuclear effects for some magnetic
moments calculated, the laboratory investigation can give some
limits for a variation of $m_e/m_p$ and $g_n$. A precision
laboratory study of transitions which are not perturbed by nuclear
effects, can give a limit for the fine structure constant and for
the ratio $g_p\,m_e/m_p\,\mu_{\sc n}$. The neutron magnetic moment
cannot be studied successfully in this way, because of the
relatively low accuracy of the deutron {\em hfs\/} and the
influence of the nuclear effects on most magnetic moments of
heavier nuclei. The proton magnetic moment can only be
investigated by using the {\em hfs\/} of atomic hydrogen,
otherwise the nuclear effects are involved. We hope that a
measurement of the hyperfine separation in the hydrogen atom will
be performed and that a reliable limit on the possible variation
of the fundamental constant will be reached.

\section*{Acknowledgements}

The idea of studying the fine structure in order to search for a
variation of the fine structure constant was learnt by me from
Klaus Jungmann \cite{jungmann} and all related possibilities arise
from discussions with him. Opportunity, provided with the hydrogen
hyperfine structure have been frequently discussed with Ted
H\"ansch. Possibilities of studying the fine structure of
different neutral atoms were discussed with Martin Weitz, while
for ions the problem was discussed with Ekkehard Peik. I was happy
to learn from them that some search for the fine structure
associated with the ground state suggested in the paper can be
really performed. I am very grateful to all of them. A preliminary
version of this work was presented at a seminar at the PTB and I
thank A. Bauch, B. Fisher and F. Riehle for comments and useful
discussions. I would like also to thank A. Godone, R. Holzwarth,
%N. Ramsey,
J. Reichert, M. Plimmer and D. A. Varshalovich for stimulating
discussions and useful references. I am thankful to Sile Nic
Chormaic for vastly improving the legibility of this paper. The
work was supported in part by the Russian State Program
`Fundamental Metrology' and NATO grant CRG 960003.

\newpage

\appendix

\section{Precision measurements of the hyperfine separation in
the hydrogen atom}

Here we give a brief description of the most important, accurate
experiments on the {\em hfs\/} as presented in Table \ref{Thhfs}.
\begin{itemize}
\item Value \#2 \cite{hellwig} is the result of an experiment
done by NBS and Harvard university. The wall-shift was measured
in early 1969 \cite{zitzewitz} and the frequency at the end of
1969.
 \begin{itemize}
 \item The result is 1 420 405.751 769 1(24) kHz. Three sources of
uncertainty were considered important.
 \item The most significant of these is the wall-shift, because the
properties of the bulb could vary with time: $\delta= 1.9\cdot
10^{-6}$ kHz.
 \item The accuracy capability
of the NBS cesium standard gave $\delta= 0.7\cdot 10^{-6}$ kHz.
 \item The frequency dispersion of the standard since it was calibrated was obtained as
 $\delta= 1.3\cdot 10^{-6}$ kHz.
\end{itemize}
\item Value \#3 (another result from the same paper
\cite{hellwig}) had a different budget of errors. This is a pure
NBS experiment, with simultaneous measurements of the wall-shift
and the frequency.
 \begin{itemize}
 \item The result for the frequency was found to be 1 420 405.751
766 7(18) kHz. The uncertainty due to the cesium standard was the
same:
 \item the accuracy capability of the standard led to $\delta= 0.7\cdot 10^{-6}$ kHz;
 \item The frequency dispersion since the calibration was
 $\delta= 1.3\cdot 10^{-6}$ kHz.
 \item The largest contribution to the uncertainty other than those
mentioned was due to a comparison of the maser to the cesium
standard: $\delta= 1.0\cdot 10^{-6}$ kHz.
\end{itemize}
\item Value \#6 \cite{morris} was an NRC experiment.
 \begin{itemize}
 \item The result was 1 420 405.751 770(3) kHz.
 \item The uncertainty of the comparison properly was only $\delta= 0.7\cdot 10^{-6}$ kHz.
 \item The variation of the maser frequency during the experiment led to
$\delta= 2.1 \cdot 10^{-7}$ kHz.
 \item The instability of the cesium standard frequency gave $\delta= 1.4\cdot 10^{-6}$ kHz.
\end{itemize}
\item An NPL experimental result (\#5) has been frequently quoted as the most
precise one \cite{essen}. However, the final NPL value (\#7) has
not been the best one.
 \begin{itemize}
 \item The result is 1 420 405.751 766 2(30) kHz and there are three sources of
 uncertainties.
 \item The measurement itself gave a statistical error $\delta= 1.4\cdot 10^{-6}$ kHz.
 \item The accuracy of the cesium standard itself gave $\delta= 1.4\cdot 10^{-6}$ kHz.
 \item Due to the same problem a comparison of the NPL cesium standard with
the international time cannot be perfect and the uncertainty was
$\delta= 1.4\cdot 10^{-6}$ kHz.
\end{itemize}
\item The result (\#8) of an experiment in Ref. \cite{reinhard} has been never
published in a refereed journal to the best of our knowledge, so
we cannot consider this as a final result. However, it is
important because they used a very different approach. In contrast
to the other experiments the wall-shift was measured by using of
a flexible bulb.
 \begin{itemize}
 \item The result was found to be 1 420 405.751 768 0(20) kHz.
 \item The maser uncertainty was $\delta= 1.4\cdot 10^{-6}$ kHz.
 \item The comparison itself was a little more accurate: $\delta= 1.2\cdot 10^{-6}$ kHz.
 \item The uncertainty due to the wall-shift was estimated as
$\delta= 0.4\cdot 10^{-6}$ kHz.
\end{itemize}
\item Result \#11 (1 420 405.751 768(2) kHz)
is an average over 5 hydrogen masers with a dispersion of their
frequencies between 1 420 405.751 765 9(17) kHz and 1 420 405.751 769 3(16) kHz.
The budgets of errors for particular masers have not been presented.
\item Result \#13 (1 420 405.751 773(1) kHz)
is not in fair agreement with most other precise values. The
method was also quite different from others as well. First, a new
maser was a developed using a flexible bulb. Next, the dependence
of the wall-shift on the temperature was studied and it was found
that the wall-shift vanished at a temperature $T_0$. The result
was found by interpolating the maser frequency to $T_0$.
\item We do not consider result \#10 (1 420 405.751 771(6) kHz)
in detail, because the accuracy is slightly lower than for
forementioned results and the statistical weight for any
evaluations is small enough. The uncertainty of a measured value
of 1 420 405.751 771 0(58) kHz was mainly due to the comparisons
of the maser frequency to the local rubidium standard and the
standard to the international time scale.

\end{itemize}

\end{document}